\documentclass[useAMS,usenatbib]{mn2e}
\def\stis{{\rm{STIS}}~}
\def\wfc{{\rm{WFC3}}~} 
\def\hst{{\it{HST}}} 
\newcommand{\bjdtdb}{\ensuremath{\rm {BJD_{TDB}}}}
\newcommand{\feh}{\ensuremath{\left[{\rm Fe}/{\rm H}\right]}}
\newcommand{\teff}{\ensuremath{T_{\rm eff}}}

\newcommand{\msun}{\ensuremath{{\rm M}_\odot}}
\newcommand{\rsun}{\ensuremath{{\rm R}_\odot}}
\newcommand{\lsun}{\ensuremath{{\rm L}_\odot}}
\newcommand{\mj}{\ensuremath{{\rm M}_{\rm J}}}
\newcommand{\rj}{\ensuremath{{\rm R}_{\rm J}}}
\newcommand{\fave}{\langle F \rangle}
\newcommand{\fluxcgs}{10$^9$ $erg s^{-1} cm^{-2}$}

\def\hat{\mbox{HAT-P-1}}
\def\tp{\mbox{\it{(T-P)}}~}
\def\hdtwo{\mbox{HD~209458b}} 
\def\hdone{\mbox{HD~189733b}} 

\usepackage{graphicx}
\usepackage{natbib}
\usepackage{epstopdf}
\usepackage{subfigure}
\usepackage{verbatim}
\usepackage{tablefootnote}
\usepackage{hyperref}
\title[HST transmission spectroscopy of HAT-P-1b with STIS]
{HST hot Jupiter Transmission Spectral Survey: A detection of Na and strong optical absorption in \hat b}
\author[N. Nikolov et al.]
 {N.~Nikolov,$^1$\thanks{E-mail: nikolay@astro.ex.ac.uk (NN)}
  D.~K.~Sing,$^1$
  F.~Pont,$^1$
  A.~S.~Burrows,$^2$
  J.~J.~Fortney,$^3$  
  G.~E.~Ballester,$^4$
  \newauthor 
  T.~M.~Evans,$^5$
  C.~M.~Huitson,$^1$
  H.~R.~Wakeford,$^1$    
  P.~A.~Wilson,$^1$
  S.~Aigrain,$^5$
    \newauthor  
  D.~Deming,$^6$    
  N.~P.~Gibson,$^7$   
  G.~W.~Henry,$^8$
  H.~Knutson,$^9$
  A.~Lecavelier des Etangs,$^{10}$ 
  \newauthor 
  A.~P.~Showman,$^4$
  A.~Vidal-Madjar,$^{10}$
  K.~Zahnle$^{11}$\\
  $^{1}$Astrophysics Group, School of Physics, University of Exeter, Stocker Road, Exeter EX4 4QL, UK\\
  $^{2}$Department of Astrophysical Sciences, Peyton Hall, Princeton University, Princeton, NJ 08544, USA\\ 
  $^{3}$Department of Astronomy and Astrophysics, University of California, Santa Cruz, CA 95064, USA\\ 
  $^{4}$Lunar and Planetary Laboratory, University of Arizona, Tucson, Arizona 85721, USA\\ 
  $^{5}$Department of Physics, University of Oxford, Denys Wilkinson Building, Keble Road, Oxford OX1 3RH, UK\\ 
  $^{6}$Department of Astronomy, University of Maryland, College Park, MD 20742 USA\\ 
  $^{7}$European Southern Observatory, Karl-Schwarzschild-Str. 2, 85748 Garching bei M{\"u}nchen, Germany\\ 
  $^{8}$Tennessee State University, 3500 John A. Merritt Blvd., P.O. Box 9501, Nashville, TN 37209, USA\\ 
  $^{9}$Division of Geological and Planetary Sciences, California Institute of Technology, Pasadena, CA 91125 USA\\
  $^{10}$CNRS, Institut dÕAstrophysique de Paris, UMR 7095, 98bis boulevard Arago, 75014 Paris, France\\
  $^{11}$NASA Ames Research Center, Moffett Field, CA 94035, USA}

\begin{document}

\date{Accepted 2013 September 30.  Received 2013 September 30; in original form 2013 July 10}

\pagerange{\pageref{firstpage}--\pageref{lastpage}} \pubyear{2013}

\maketitle

\label{firstpage}

\begin{abstract}

We present an optical to near-infrared transmission spectrum of the hot Jupiter \hat b, based on {\it{HST}} observations, covering the spectral regime from  $0.29$~to~$1.027\mu m$ with STIS, which is coupled with a recent {{\wfc}} transit (1.087 to 1.687$\mu m$). We derive refined physical parameters of the \hat~system, including an improved orbital ephemeris. The transmission spectrum shows a strong absorption signature shortward of $0.55\mu m$, with a strong blueward slope into the near-ultraviolet. We detect atmospheric sodium absorption at a $3.3\sigma$ significance level, but find no evidence for the potassium feature. The red data implies a marginally flat spectrum with a tentative absorption enhancement at wavelength longer than $\sim0.85\mu m$. The \stis and \wfc spectra differ significantly in absolute radius level ($4.3\pm1.6$ pressure scale heights), implying strong optical absorption in the atmosphere of  \hat b. The optical to near-infrared difference cannot be explained by stellar activity, as simultaneous stellar activity monitoring of the G0V \hat b host star and its identical companion show no significant activity that could explain the result. We compare the complete \stis and \wfc transmission spectrum with theoretical atmospheric models which include haze, sodium and an extra optical absorber. We find that both an optical absorber and a super-solar sodium to water abundance ratio might be a scenario explaining the \hat b observations. Our results suggest that strong optical absorbers may be a dominant atmospheric feature in some hot Jupiter exoplanets.

\end{abstract}


\section{Introduction}
Since the first detection of an exoplanet hosted by a Sun-like star \citep{mayor95}, the number of extrasolar planets and candidates has rapidly grown, yet exceeding more than 3000. Among them, the planets that temporarily pass in front (transit) or behind (secondary eclipse) the host star offer a unique opportunity to explore their dynamical properties, internal structures and atmospheres \citep{winn10a,seager11a}. Specifically, during transits part of the star light travels through the rarified upper planetary atmosphere (near the terminator) and is partially absorbed by atoms and molecules. This process gives rise to a wavelength dependent variation of the measured transit depth and is central to the method of transmission spectroscopy, the subject of this paper.  The optical transmission spectrum of a cloud free hot Jupiter exoplanet at temperatures lower than $1500\,K$ is predicted to be dominated by pressure broadened sodium (Na\,I) and potassium (K\,I) resonance lines \citep{brown01, seager00, burrows10, fortney10}. 

Currently, observations have led to mixed detections of both strong alkali features in several exoplanets with \hdtwo~and \hdone~remaining among the two most studied exoplanets, due to the prominent brightness of their host stars ($V\sim8$ mag) and deep transit signals. In particular, the optical transmission spectrum of \hdtwo~shows broad sodium feature and a lack of potassium \citep{charbonneau02, sing08b, snellen08, narita05}. The complete optical to near-infrared spectrum of \hdone~is described by an increasing planetary radius with decreasing wavelength, most likely due to Rayleigh scattering \citep{pont08, sing09b, sing11b, lecavelier08} and presence of narrow line cores of sodium peaking above the slope \citep{redfield08, huitson12, jensen12}. Although there is no reason to expect that these two atmospheres should be similar, as infrared observations indicate that the atmosphere of \hdtwo~has a temperature inversion \citep{knutson08a, burrows07b} while the atmosphere of \hdone~does not \citep{grillmair07, charbonneau08}, the significant difference between both planetary atmospheres indicates that clouds and hazes can be very important in determining the overall shape of transmission spectra. In addition to \hdtwo~and \hdone, a sodium core has also been detected in WASP-17b but with an absorption depth much less than the theoretical prediction \citep{wood11, zhou12}. \cite{wood11} speculate that the depleted sodium abundance could be due to photoionization. Notably, XO-2b is the first exoplanet for which both sodium and potassium absorption was detected \citep{sing11a, sing12}. Unlike \hdone, this planet shows no evidence for {\rm{high-altitude}} clouds or hazes.

\begin{figure*}
\includegraphics[width=\textwidth,height=\textheight,keepaspectratio]{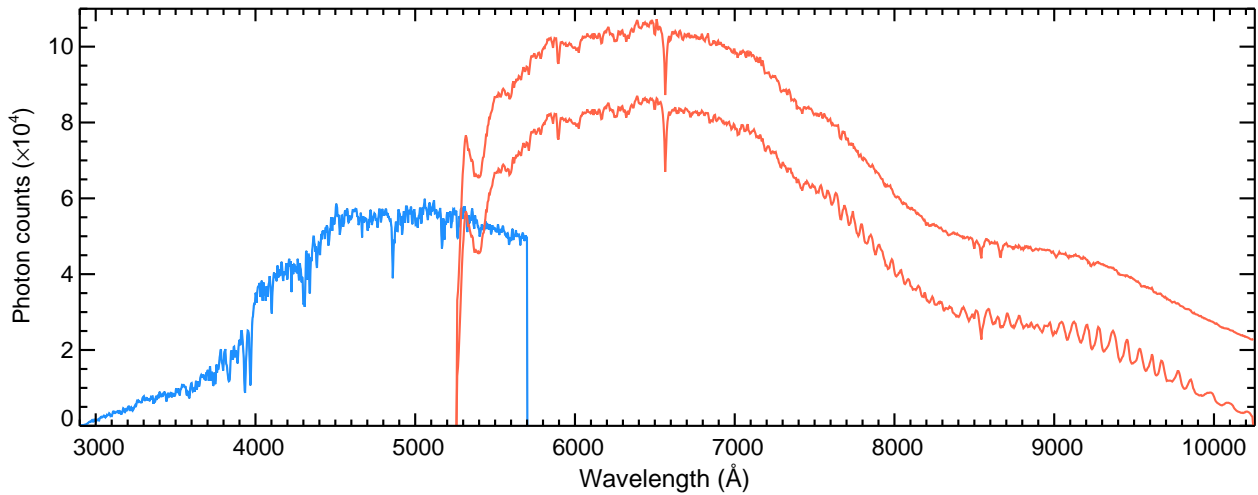}
\caption{Typical one-dimensional G430L (blue line) and G750L (red line) spectra of \hat extracted from our time series. 
Optical element G430L covers the wavelength range from 2892 {\AA} to  5700 {\AA}, while
G750L spans the range from 5240~{\AA} to 10270~{\AA}, enabling a construction of the complete
optical to near-infrared transmission spectrum of extrasolar planets at low-resolution. 
Note the strong long wavelength fringing beyond 
$\sim7000$ {\AA} in grating G750L (displayed on the lower red spectrum) compared to 
a fringe-corrected data (displayed in the upper red spectrum shifted by $2\times10^{4}$ counts for clarity). A colour version is available in the online journal.}
\label{fig:stis_spectrum}
\end{figure*}

Motivated by the recent observational results and theoretical predictions of strong alkali absorption features, stratosphere inducing TiO and sulphur bearing molecular features, near-infrared molecular features, clouds, hazes and dust we initiated an optical to near-infrared spectroscopic survey of eight close-in giant exoplanets across different regimes of planetary temperature (in the range $1080-2800\,K$). The ultimate aim of the programme is to classify the monitored targets based on their atmospheric specificity, i.e. clear/hazy, presence/lack of TiO/alkali and other molecular features. Accumulating a large number of observed planets will open the possibility to measure correlations between atmospheric properties and other parameters such as stellar activity and stellar type. In this paper we report results from three new transit observations of \hat b with the G430L and G750L gratings of the {\rm{Space Telescope Imaging Spectrograph (STIS)}} aboard the {\it Hubble Space Telescope (HST)}. In addition, we complement our spectroscopic data with the recently reported \hst/{\rm{Wide~Field~Camera~3 (WFC3) G141}} transit of \cite{wakeford13} and investigate the complete optical to near-infrared transmission spectrum of \hat b with theory of giant exoplanet atmospheres. 

This paper is organised as follows: Section~\ref{sec:observationsec}  presents the instrument set-up employed for the transit observations and the data reduction. Section~\ref{sec:analysis} describes the data analysis, including a refinement of the system parameters and the construction of the transmission spectrum of \hat b. Finally, we discuss the results and conclude in Sections~\ref{sec:discussionsec} and~\ref{sec:conclusionsec}.

\subsection{The hot Jupiter \protect\hat b}

Detected by the small automated telescopes of the HATNet ground-based transit survey, \citep{bakos07} \hat b is a giant ($R_{{\rm{p}}}~\sim~1.2~R_{{\rm{J}}}$), low-mean density ($\rho_{{\rm{p}}}~\sim~0.4~{{\rm{g  cm^{-3}}}}$) transiting extrasolar planet in a visual binary system, composed of two sun-like stars (separated by angular distance of $11\arcsec$ on the sky, corresponding to a linear distance of $\sim1500$ AU) located more than 450 light-years away in the northern constellation Lacerta. The host star (\hat~also known as ADS~16402~A, BD+37~4734, $\alpha=22^{\rm{h}}57^{\rm{m}}46.8^{\rm{s}}$,  $\delta =+38^{{\rm o}} 40^{\prime}28^{\prime\prime}$, J2000.0) is a moderately bright G0V dwarf ($V=10.4$ mag) allowing for characterisation follow-up studies. Although it has been suggested by \cite{bakos07}  that \hat b is too large to be explained by theoretical models of giant exoplanets, \cite{winn07} improved the accuracy of the system parameters of \hat~and its planet and concluded that the planet radius is in accord with theoretical models of irradiated, coreless, solar-composition giant planets. Furthermore, \cite{johnson08} combined {\it{Keck I/{\rm{HIRES}}}}\footnote{High Resolution Echelle Spectrometer} and {\it{Subaru/{\rm{HDS}}}}\footnote{High Dispersion Spectrograph} optical spectra with photometry, detecting the Rossiter-McLaughlin effect and measuring a close alignment between the  planet's orbital inclination and the rotation axis of the star $\lambda = 3.7^{\circ}\pm2.1^{\circ}$. \cite{liu08} reexamined the core mass needed for \protect{\hat} b, compared to the available measurements of its radius at the time and concluded that their inferred core masses are roughly consistent with the stellar metallicity versus core-mass relationship. 

The atmosphere of the planet has also been probed with secondary eclipse observations. \cite{todorov10} reported {\it{Spitzer/{\rm{IRAC}}}}\footnote{Infrared Array Camera}  photometry of \hat b during two secondary eclipse observations,  covering broad-bands at $3.6\mu$m, $4.5\mu$m, $5.8\mu$m and $8\mu$m. The authors found best-fits for their results of the secondary eclipse depths using an atmosphere with a modest temperature inversion, intermediate between the archetype inverted atmosphere of \hdtwo~and a model without an inversion. Further in their discussion the authors speculated that the best fit for the average dayside temperature of \hat b is $1500\pm100\,K$. Analysis of the secondary eclipse led the authors to the conclusion that the orbit of \hat b is close to circular, with a $3\sigma$ limit of $|e \cos{\omega}| < 0.002$. Recently, \cite{beky13} reported visible-light {\hst}{/}{\stis} relative photometry during two occultations of \hat b, allowing them to constrain the geometric albedo of the planet. Comparing two techniques: (i) relative photometry (of \hat\,A with respect to its companion \hat\,B) and (ii) the traditional steps in removing instrumental artefacts from (single target) \hst~time-series, the authors concluded that the second method introduced a strong bias in the albedo result. The authors estimated a $2\sigma$ upper limit of 0.64 for the geometric albedo of \hat b between 5770 and 9470 \AA.

\section{Observations and Calibrations}\label{sec:observationsec}

We obtained low-resolution optical to near-infrared spectra of \hat~with the STIS instrument aboard the {\it{HST}} (Proposal ID GO-12473, P.I., D. Sing) 
during three transits on UT~2012~May~26 and 30 with grating G430L and 2012 September~19 with grating G750L. 
Table~\ref{tab:obsdatetab} exhibits a summary of our observations. 

Each visit consisted of five $\sim96\ {\rm{min}}$ orbits, during which data collection
was truncated with $\sim45\ {\rm{min}}$ gaps due to Earth occultations. 
Given the moderate brightness of \hat~($V=10.4$ mag) the integration time was set to
$284\ {\rm{s}}$ to achieve a high signal-to-noise ratio, resulting in 43 spectra
during each visit. 
Each of the G430L and G750L optical elements
possesses a resolving power of $R=500$, which secures a combined 
wavelength coverage from 2892~{\AA} to 10270~{\AA}
with a small overlap region between them from 5,240 {\AA} to 5,700 {\AA} (see Fig.~\ref{fig:stis_spectrum}).  
Similar to \cite{sing11a} the data for this study were collected 
with a wide $52\times2''$ slit 
to minimise slit loses. Data acquisition overheads were minimised
by reading-out a reduced portion of the CCD with a size of $1024\times128$.

\begin{figure*}
\includegraphics[width=\textwidth,height=\textheight,keepaspectratio]{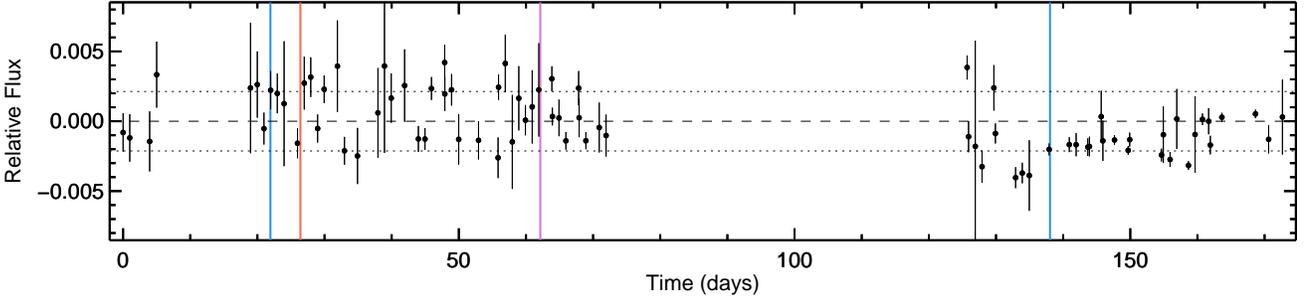}
\caption{\protect{\hat}b stellar variability monitoring performed with the 2.5\,m {\it{Liverpool Telescope}} spanning 223\,days starting UT~2012~May~4. {\hst}/{\stis}\,G430L, G750L and {\hst}/{\wfc}\,G141 visits are indicated with blue, red and purple vertical lines, respectively. The dashed and the two dotted lines indicate the mean and the standard deviation values of the stellar flux.  A colour version is available in the online journal.}
\label{fig:variability_plot}
\end{figure*}

Each of the three visits consisted of five orbits during which nine spectra were collected. Three of the orbits at each visit were scheduled to provide out-of transit measurements to secure a stable baseline flux while two orbits occurred during the transit events. Therefore, when combined the three visits provide an almost complete light curve, needed for an accurate measurement of the system parameters and the planet radius.




\begin{table}
\centering
\caption{{\it{HST}} STIS observing dates, instruments and settings.}
\begin{tabular}{@{}ccccc}
\hline
UT date & Visit & Optical & Number & Integration  \\
 2012     &        \#      & element & of spectra & time (s) \\
\hline
May 26             &   7    & G430L & 43 & 284\\
May 30             &  20  & G750L &  43  & 284\\
September 19 &   8   & G430L &  43  & 284\\
\hline
\label{tab:obsdatetab}
\end{tabular}
\end{table}

The data was reduced 
(bias-, dark- and flat-corrected) using the 
latest version of the {\tt{CALSTIS}}\footnote{{\tt{CALSTIS}} comprises software tools developed for the 
calibration of STIS data \citep{katsanis98} inside the {\tt{IRAF}} (Image 
Reduction and Analysis Facility; http://iraf.noao.edu/) environment.} pipeline and 
the relevant up-to-date calibration frames. 

{\rm{STIS}} observations with the G750L grating  
are prone to significant fringing 
at wavelengths longward of $\sim7000$\,\AA, which can limit the signal-to-noise ratio ({\sl{SNR}}) of the derived spectra
 \citep{goudfrooij98a}.
The fringes are caused by interference of multiple reflections between the two surfaces 
of the thinned, backside-illuminated SITe CCD onboard STIS.
Although the fringe pattern may have a negligible 
effect on transmission spectra, acquired from
broad spectral widths ($\sim1000$\,\AA), as pointed out
by \cite{knutson07} in their analysis of \hdtwo, the effect may be 
significant at smaller band widths and requires a careful 
treatment of the fringe pattern on each of the G750L science spectra. 

We corrected the fringe pattern present on each of the G750L 
science spectra of the time series, using a contemporaneous fringe flat image (FFI) and 
the recipe of \cite{goudfrooij98b}.
To summarise, we removed the low order lamp vignetting 
and normalised the FFI. We then aligned and scaled the 
infrared fringes to match the fringe pattern of 
the star spectrum on each of the science spectra.
Finally, the stellar spectrum was divided by the shifted flat, to 
correct for the fringe pattern (see Fig.~\ref{fig:stis_spectrum} for a comparison
between an original G750L spectrum, exhibiting fringe effect 
to its defringed version).

\subsection{Cosmic ray correction}

An examination of the spectral time series showed a high number of recorded cosmic ray (CR) events. This is not unusual given the relatively long integration time of $284\ {\rm{s}}$, employed for our study on \hat b. For comparison, previous studies on transiting exoplanets with STIS in a low-resolution mode, such as \cite{knutson07} on HD~209458 ({\rm{V=7.65}}) and \cite{sing11b} on ${\rm{HD\ 189733}}$ (V~$=$~7.67 mag), used integration times of $\sim20\ {\rm{s}}$ and $\sim64\ {\rm{s}}$, respectively. The application of such shorter integration times (i.e. factor of $\sim14$ and $\sim4$ for the afore-mentioned studies compared to ours) on bright target stars serves to secure time series of spectra with a lower number of CR events and good sampling of the resulting light curves. When recorded on time series of two-dimensional spectra, the CRs act to effectively increase (though ``artificially") the flux of the targeted star on random locations within a limited area along the spectra, and appear as outlier data points on the light curves originating from the affected wavelength bins. We therefore consider the CRs present in our data as a potential source of light curve systematics and undertook a careful analysis aiming CR identification and correction on the science two-dimensional spectra.

Initially, we aimed to perform the data analysis 
utilising the {\tt{.crg}} images
produced after a CR identification and 
correction by {\tt{CALSTIS}}. However, 
it was found in the subsequent analysis 
that for most of the spectra 
the CR identification was inadequate.
In particular, large portions or in several cases complete 
lines around the central regions of
the spectra were found to be marked as 
affected by CR events lacking a justification
when compared to the original spectrum.
We therefore developed 
an alternative customised method 
to i) perform a more efficient 
CR identification and ii) to correct the affected regions of the spectra.  
To identify the CRs on each image, 
we inspected difference images, constructed from the elements of the time series.  
First we computed the differences between each 
image we aimed to correct for CRs and four neighbouring images:
two consecutive images obtained before and two after the image we aim to correct. 
In this process, each of the resulting four difference images were dominated by 
a three-component signal. 
The first and the second major components
were determined by the CR events specific 
for the neighbour image itself and the CRs that were present on 
the image we aim to correct. The third component
was a negligible residual of
the stellar spectrum
due to variations associated with the
observed target (e.g. transits or other short-time scale events) 
and/or instrumental systematics (e.g. temperature variations of the telescope).
We eliminated the 
first component, and dramatically reduced the amplitude of the 
third, after a computation of 
the median of the four difference images,
which resulted in an image containing the CR events only.
The later property of the median-combined image
allowed us to identify CRs even at locations quite 
close to the peak of the stellar point spread function (PSF).

To identify the pixels affected by CRs on each 
median-combined image, 
we scanned each of the CCD's 128 lines, using a window 
centred on the pixel we were inspecting. 
{\rm{The window had a size of 20 pixels,
determined by the typical size of a CR on the detector.}} 
We compared the value of the 
central pixel in the window 
with the median value of all pixels within 
the window and flagged central pixels that were 
found to be 5$\sigma$ above the median level (where $\sigma$ is the standard deviation of the examined line pixel values). 
To flag huge regions of the spectra 
affected by CRs (i.e. regions with size exceeding the window size), we also compared the 
central pixel values with the median 
value of the entire line and again 
flagged pixels that showed 
values higher than 5$\sigma$ 
above the median level for that line. 
Next, we used all flagged pixels and computed a nominal 
PSF profile for each column, using five neighbour columns (ignoring the CR affected pixels in them) 
before and after the examined column. We then scaled the nominal column to match the one with affected pixels. 
Finally, all flagged pixels were replaced with their corresponding values in the scaled profile.
Following this procedure we found that the total number of pixels affected by CR events
comprise $\sim5\%$ of the total number of pixels on the science images. In addition, we also
substituted all pixels identified by {\tt{CALSTIS}} as ``bad" with
the same procedure.

\begin{figure*}
\includegraphics[width=\textwidth,height=\textheight,keepaspectratio]{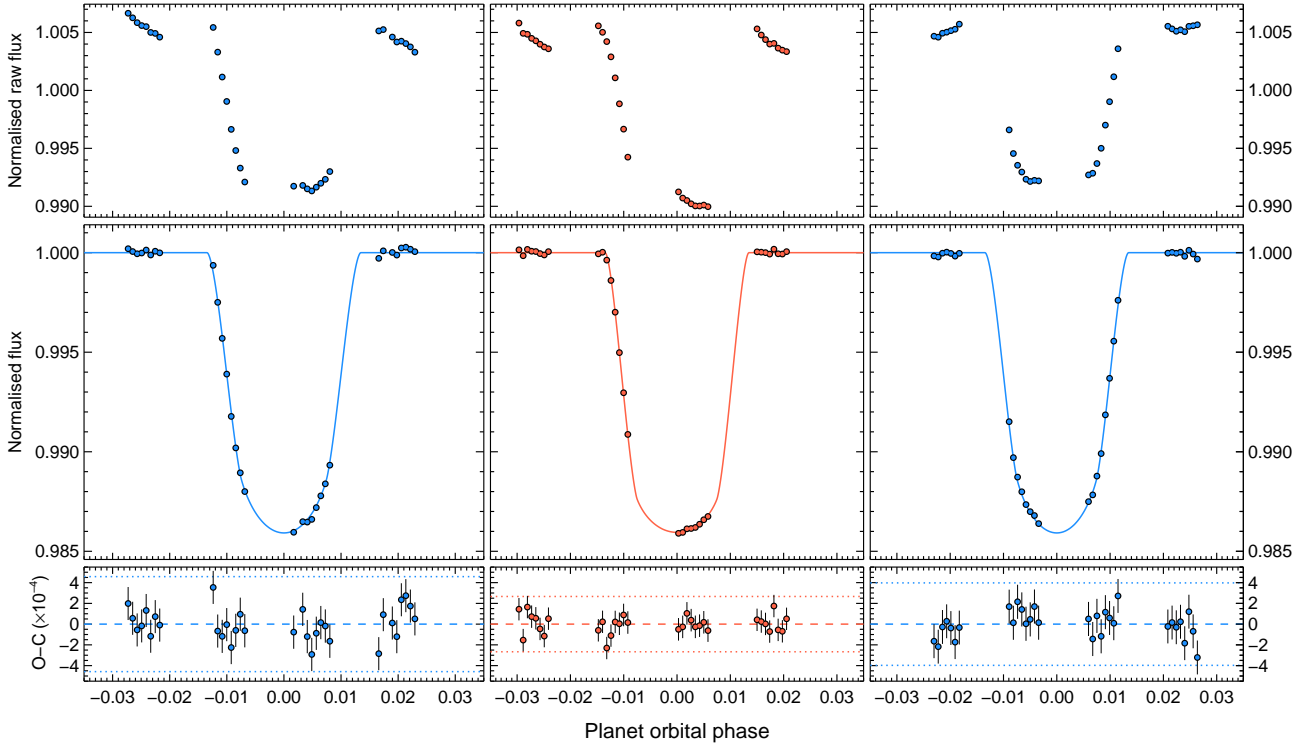}
\caption{ {\it{HST}}/STIS normalised white light curves
 based on data collected during the three visits (left to right):
 on UT 2012 May 26 (G430L), May 30 (G750L) and
September 19 (G430L). 
{\sl{Top panels:}} Raw light curves normalised to
the mean raw flux (originally in electrons). The light curves experience prominent systematics associated with the {\it{HST}} thermal cycle (see text for details); 
{\sl{Middle panels:}} Detrended 
light curves along with the best-fit transit model \protect\citep{mandel02} superimposed 
with continuous lines; 
{\sl{Lower panels:}} Observed minus modelled light curve 
residuals, compared to a null (dashed lines) and a $3\sigma$ level (dotted lines)
used to identify outliers. The spectrophotometric data from G430L and G750L are colour coded in blue and red, respectively.
A colour version is available in the online version of the journal.}
\label{fig:wlc_stis}
\end{figure*}

\subsection{Stellar variability monitoring}\label{sec:stellaractivity}

We used the RISE\footnote{{\bf{R}}apid {\bf{I}}maging {\bf{S}}earch for {\bf{E}}xoplanets, \cite{steele08}} camera on the 2.5\,m {\it{Liverpool Telescope (LT)}} at the Roque de los Muchachos Observatory in La Palma, Canary Islands to monitor the brightness of \hat~for evidence of stellar variability (see\,Fig.~\ref{fig:variability_plot}). Commencing on UT\,2012\,May\,4 and ending on UT\,2012~December~13, the observations spanned a period of 223 days that included the three STIS visits and the WFC3 visit. A series of 300 exposures with integration times of 0.8\,s were taken each
night with $2 \times 2$ pixel binning\footnote{Such short integration times are only possible due to the short read-out time of RISE.}. The telescope was slightly defocussed to prevent saturation of the detector and to spread the light across a larger number of pixels and minimise flat fielding errors.

Standard {\tt{IRAF}}\footnote{The acronym {\tt{IRAF}} stands for Image Reduction and Analysis Facility.}/{\tt{PHOT}} routines were used to obtain accurate centres and perform aperture photometry for \hat~and three other bright stars within the RISE camera's $9.2^\prime \times 9.2^\prime$ field of view. The defocussed PSFs typically had full widths at half maximum of $10-12$\,pixels ($5.4^{\prime\prime}
-6.5^{\prime\prime}$), so we adopted a radius of 9 pixels for the photometric aperture. Larger aperture radii were not possible, due {\rm{to}} the nearby companion to \hat~separated by only $\sim\,20$\,pixels ($11^{\prime\prime}$). The sky contribution was estimated by taking the median pixel count within an annulus centred on each star, with an inner edge radius of 30 pixels and a width of 60 pixels.

To correct for night-to-night changes in telluric conditions and the instrumental setup, we used the comparison stars to perform differential photometry for \hat. We experimented with dividing the raw fluxes for \hat~by different combinations of the comparison stars' raw fluxes. In the end, however, we only used the nearby bright comparison star, as the others were simply too faint.

\subsection{\stis white light curves}

Spectral extractions were performed in {\tt{IRAF}} using 
the {\tt{APALL}} procedure. The photon flux originating
from a wavelength range
in each spectrum of the time series
was summed to produce a light curve for that wavelength bin. Photometric per-point uncertainties were initially derived based on pure photon statistics.
Similar to other spectrophotometric
studies, we refer to the
light curve computed after a summation of 
the complete wavelength range as ``white" light curve.

An optimum performance of the spectral extraction procedure
was secured by a selection of the combination of an aperture size
and background subtraction
(among various aperture sizes)
that minimise the out of transit (oot) root-mean square (rms) flux
of the resulting white light curve. It was found that 
the smallest scatter was achieved
with a 13 pixel wide aperture.
A wavelength solution was obtained from the {\tt{\_x1d}} files from {\tt{CALSTIS}}. 
Similar to previous {\it{HST}}/STIS studies on transiting exoplanets,
the first orbit of each transit 
observation (purposely scheduled earlier than each transit event)
as well as the first exposure of each orbit were discarded.
That data exhibits complex and unique 
systematics, as it was 
acquired while the telescope was thermally relaxing into its 
new pointing position.
Following the afore-mentioned data quality constraints, 
each of the
three STIS transit observations resulted in 
four orbits, each of which containing eight
measurements (see the top panels of Fig.~\ref{fig:wlc_stis}).

\subsection{A WFC3 near-infrared complement}

\begin{figure}
\includegraphics[width=0.5\textwidth,height=\textheight,keepaspectratio]{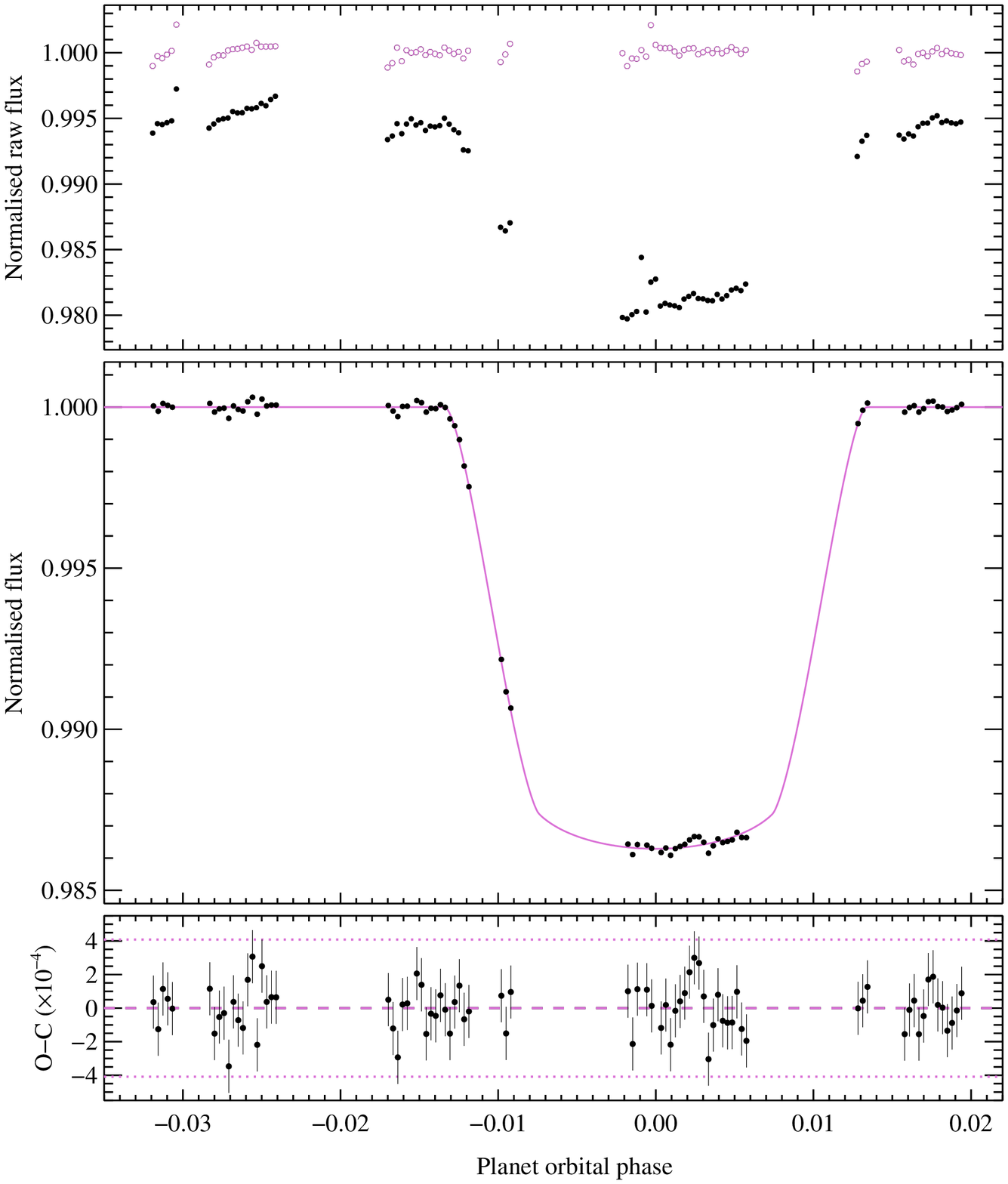}
\caption{ {\it{HST}}/WFC3 transit data from \protect\cite{wakeford13}, included in the simultaneous fit. {\sl{Top panel:}} Raw white light curves of a comparison star (open circles in purple) and \protect{\hat} (black dots); {\sl{Middle panel:}} Relative photometry of \protect{\hat} with respect to the comparison star along with the best-fit transit model \protect\citep{mandel02} displayed with a purple line; {\sl{Lower panel:}} Observed minus computed residuals. The three sigma residual rms levels are indicated with dotted lines compared to zero level. A colour version is available in the online version of the journal.}
\label{fig:wlc_wfc3}
\end{figure}

Results for \hat b from an {\it{HST}}/WFC3~G141 transit observation (Visit 26) in drift scanning mode have been reported by \cite{wakeford13}. The data comprised of four orbits, each containing $\sim$~45 min intervals of data collection with an integration time of $\sim$~47 sec. Notably, the WFC3 field of view also permitted data sampling of the PSF of star HAT-P-1B, which is the second component of the G0V/G0V visual binary that contains the host of \hat b. Since HAT-P-1B is low active star, it may serve as a comparison star for relative photometry. The white light curve based on the {\it{HST}}/WFC3 data was also useful in determining the system parameters of \hat. We therefore include that light curve in the analysis of the STIS data (see the top panels of Fig.~\ref{fig:wlc_wfc3}).

\section{Analysis}\label{sec:analysis}
A transit light curve analysis
was performed for each visit
employing a two-component
model fit to the data.
The first component is 
based on the complete analytic transit formula given in 
\cite{mandel02}, which in addition to the central transit times
($T_{{\rm{C}}}$) and orbital period ($P$), is a function of the orbital inclination
($i$), normalised planet semi-major axis ($a/R_{\ast}$) 
and planet to star radius ratio ($R_{{\rm{p}}}/R_{\ast}$). 
To account for the 
limb darkening of \hat~we, employ the 
four parameter non-linear limb darkening law 
defined in \cite{claret00} given by,


\begin{equation}
 \frac{I(\mu)}{I(1)} = 1 - \sum_{n=1}^{4}  c_{n}(1-\mu^{n/2})
\end{equation}
where $I(1)$ is the intensity at the centre of the stellar disk
 and $\mu=\cos{\theta}$, where 
$\theta$ is the angle between the line of sight and 
the normal to the stellar surface and $c_n, n=1,4$ are the four limb darkening coefficients.  

We choose to rely on theoretically derived stellar limb darkening coefficients comparing the data to both {\tt{1D}} and {\tt{3D}} stellar models, rather than to fit for them in the data in order to reduce the number of free parameters in the fit (typically four parameters per grating). Furthermore, this approach also eliminates the well-known wavelength-dependent degeneracy of limb darkening with transit depth \citep{sing08a}.

Initially, the values for the four limb darkening coefficients were derived from the {\tt{1D ATLAS}} theoretical stellar models of  Kurucz\footnote{\url{http://kurucz.harvard.edu/}}, following the procedures described in \cite{sing10}. In particular, we obtained theoretical limb darkening coefficients for the closest match to a star with the physical properties of \hat, i.e. $T_{{\rm{eff}}}=6000$ K, $\log{g}=4.5$ and ${\rm{[Fe/H]=0.0}}$. 
\begin{table}
\centering
\caption{Spectroscopically derived stellar atmospheric 
parameters for HD~209458 and \protect\hat.}
\begin{tabular}{@{} l c c c }
\hline
\hline
Property    &    HD~209458   &   \protect\hat \\
                       &  \cite{hayek12} & \cite{torres08} \\
\hline
$T_{ {\rm{eff}} },\,K$  &     $6095\pm53$    &  $5975\pm120$  \\
$\log{g} ,\ cm\,s^{-2}$                 &     $4.30\pm0.09$   & $4.45\pm0.15$\\
 ${\rm{[Fe/H]}}$,\ dex                &     $0.00 \pm0.04$   & $0.13\pm0.08$\\
\hline
\label{tab:limbdarktab}
\end{tabular}
\end{table}
Previous analyses on high signal-to-noise transit light curves with limb darkening coefficients derived from {\tt{1D}} model predictions sometimes resulted in poor fits, especially in the ingress and egress phases of the transit, which is characteristic of incorrect limb darkening \citep{hayek12}. The main reason for this issue lies in a generic shortcomings in the structure of {\tt{1D}} model atmospheres compared to more sophisticated {\tt{3D}} stellar atmospheric models. In particular, when compared in the case of the solar atmosphere, {\tt{3D}} models explicitly take into account the effect of convective motions in the surface granulation and reproduce the solar atmosphere with a higher degree of realism. \cite{hayek12} employed {\tt{3D}} stellar atmospheric models
and computed limb darkening coefficients for HD~209458. 
Notably, the stellar atmospheric parameters of \hat~are 
quite similar (at the $1\sigma$ level) to those
of HD~209458 as displayed in Table~\ref{tab:limbdarktab}. In particular, both stars 
are of similar effective temperature, however HD~209458 is $120\,K$ hotter than \hat. That difference however, is well within 
the effective temperature uncertainties of 
both stars, which allows one to 
adopt the available HD~209458 limb darkening coefficients 
in the analysis of \hat. We compare both the {\tt{1D}} and {\tt{3D}} models in the forthcoming analysis sections.

 Previous \stis data analyses showed that the first integration exhibits abnormally low flux \citep{charbonneau02, sing08b, pont08, sing11a, huitson12}. We attempted to resolve this issue by incorporating an additional 1\,s long exposure prior to the 284\,s integrations. However, it has been found that skipping the 1\,s and the first 284\,s integration of each orbit improved the fit by reducing the $\chi^2$ value. We therefore exclude these two data points from each orbit in the analysis. 

The raw STIS light curves exhibit instrumental systematics similar to those described by \cite{gilliland99} and \cite{brown01}. In summary, the major source of the systematics is related with the orbital motion of the telescope. In particular the {\sl{HST}} focus is known to experience  quite noticeable variations on the spacecraft orbital  time scale, which are attributed to thermal contraction/expansion (often referred as the `breathing effect') of the optical telescope assembly  as the telescope warms up during its orbital day  and cools down during orbital night \citep{hasan93, hasan94, suchkov98}. We take into account the systematics associated with the telescope temperature variations in the transit light curve analysis using a baseline function, which we multiply to the transit model in flux. Similar to past STIS studies, the main ingredient of the baseline function  is a polynomial of fourth degree of the {\sl{HST}} orbital phase ($\phi_{t}$).  In addition, we find in the light curve analysis that the systematics vary from orbit-to-orbit (i.e. dependence with time, $t$), which is a known effect from previous {\stis} studies as well as the detector positions of the spectra, as determined by the spectral trace orientation $(x,y)$ obtained  from the {\tt{APPAL}} task in {\tt{IRAF}} and the shift ($\omega$) of each spectrum of the time series, compared to a reference (i.e. cross correlated typically with the first spectrum).  Model selection was further investigated including  i) polynomials of degrees higher than fourth-order for the {\sl{HST}} orbital phase, and ii) additional terms to the planet orbital phase, spectral shifts and traces. However those models were found statistically unjustified for this particular dataset and hence rejected based on the Bayesian Information Criterion (BIC) introduced by \cite{schwarz78}:

\begin{equation}
{\rm{BIC}}=\chi^{2}+k\ln{n},
\end{equation}
where $k$ is the number of free parameters and $n$ is the number of data points. A summary of the model selection analysis is presented in Appendix A1. 

Systematics correction was also performed following the divide out of transit (divide-oot) method described in \cite{berta12}, where a template correction is constructed using the data from the out of transit orbits. Although the idea would work well for instruments with strictly  repeating systematics (such as data from {\it{HST}}/WFC3\footnotetext{Wide Field Camera 3}) we found that the resulting out of transit light curve root-mean-square for our STIS white or binned light curves is a factor of a few higher than the one from the conventional detrending method (adopted in this work). In addition the timing of the {\sl{HST}} exposures is known to change slightly from orbit-to-orbit relative to the {\sl{HST}} orbital phase, which makes the systematics associated to each orbit unique. We therefore  chose to remove the systematics using a parameterised fit to the data. 

An accurate determination of reliable uncertainties of the system parameters often requires a careful treatment of the photometric uncertainties. Time-correlated or ``red" noise  is often present in high precision photometry and can influence the final parameter determination. To investigate the noise in our light curves, we applied the Õtime-averagingÕ procedure \citep{pont06, winn09} and binned light curve residuals in time by $n$ data points. For each light curve we computed the $\beta-$ratio between the rms residual of the binned ($\sigma_{N}$) and unbinned ($\sigma_{1}$) data. In case that the data is free of red noise  it is expected that the binned (in $M$ bins) residuals follow the relation:

   \begin{equation}
         \sigma_{N}=\frac{\sigma_{1}}{\sqrt{N}}\sqrt{\frac{M}{M-1}}.
   \end{equation}
However, often in reality $\sigma_{N}~\geq~\sigma_{1}$ by the factor $\beta$. The $\beta$ factor is often used to rescale each photometric uncertainty of a data set. Since our data contains only nine data points for each orbit of the four visits in each light curve, it was a difficult task to obtain reliable estimates of the $\beta$ factor for large bin sizes ($N>15$). However, we find $\beta\sim1.3, 1.2, 1$ and $1.4$ for the white light curves of visits 7, 8, 20 and the  {\sl{WFC3}} white light curve  when binning no more than fifteen data points.

We determine the best-fit parameters of the two-component function to the data using the Levenberg-Marquardt  least-squares algorithm as implemented in the {\tt{IDL\footnote{The acronym {\tt{IDL}} stands for Interactive Data Language.} {\tt{MPFIT}}}}\footnote{\url{http://www.physics.wisc.edu/~craigm/idl/fitting.html}} package \citep{markwardt09}. The final results for the uncertainties of the fitted parameters were taken from {\tt{MPFIT}} after a rescale for any non-unity reduced $\chi^2$ values.

\begin{figure}
\includegraphics[trim = 10 0 0 0, clip, width = 0.48\textwidth]{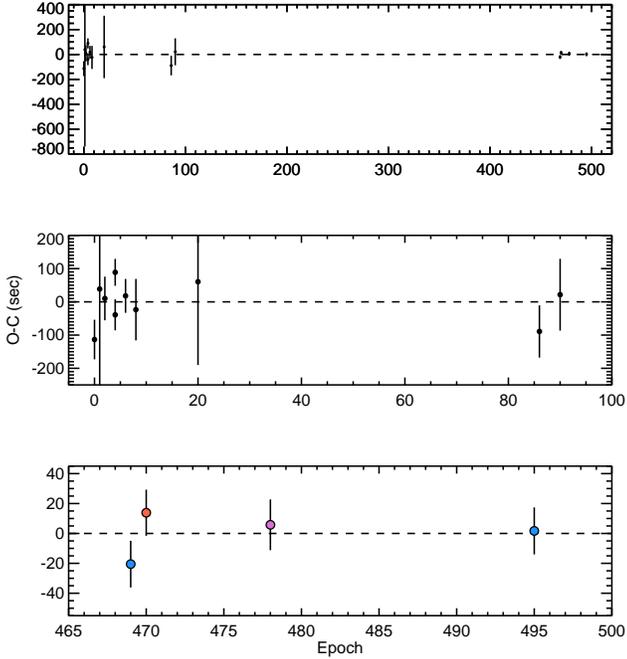}
\caption{{\sl Top panel:} Observed minus computed (O-C) diagram for \protect{\hat}~b after a combined analysis of previous transit observations with the three transits presented in this work. {\sl Middle panel:} A zoom around the first group of data points; {\sl Lower panel:} A magnification around the three central times reported in this work and the \wfc measurement of \protect\cite{wakeford13}, colour coded with blue (G430L), red (G750L) and purple (G141). A colour version is available in the online version of the journal.}
\label{fig:ocfig} 
\end{figure}

\subsection{System parameters and transit ephemeris}\label{sec:syspar}
One of the goals of the present study is to derive accurate system parameters of \hat b including an improved transit ephemeris. To achieve this task, ideally one would use a complete and well-sampled transit observation at wavelengths close to or in the infrared, where the stellar limb darkening is minimised and allows accurate parameters to be determined. In this study we rely on  the white light curves to derive system parameters. When combined, these curves span nearly a complete transit, sampled  once each 294\,s, covering the full optical to near-infrared  wavelength rage. While several complete transit light curves have been reported from the ground, we consider the quality of the STIS and WFC3 white light curves as high and the wavelength coverage wide. We therefore, choose to model the four {\sl{HST}} visits simultaneously.  
\begin{table*}
\centering
\caption{ {\tt{3D}} limb-darkening coefficients employed in the simultaneous fit and results for the fitted $R_{\rm{p}}/R_{\ast}$ and the residual scatter in parts-per-million (ppm).}
\begin{tabular}{@{} c c c c c c c c}
\hline
\hline
Visit ID &  Instrument    &  $u_{1}$ & $u_{2}$ & $u_{3}$ & $u_{4}$  & ${\rm{R_{\rm{p}}/R_{\ast}}}$ & rms (ppm) \\
\hline
7    &  {\it{STIS/G430\,L}} & 0.4397  &     0.3754     &   0.1005    &   -0.0622    &   $0.11849\pm 0.00046$& 120\\
8    &  {\it{STIS/G430\,L}} & 0.4397  &     0.3754    &    0.1005    &   -0.0622  &   $0.11849\pm 0.00046$& 133\\
20  &  {\it{STIS/G750\,L}} &0.7093   &    -0.2265    &    0.3273    &   -0.1240  &   $0.11808\pm 0.00034$& 81\\
26  &  {\it{WFC3/G141}} & 0.6958    &   -0.3332   &     0.3787    &  -0.1267  &   $0.11763\pm 0.00028$& 182\\
\hline
\label{tab:colordepthtab}
\end{tabular}
\end{table*}

Furthermore, as the orbital inclination ($i$) and the normalised semimajor axis ($a/R_{\ast}$) should not depend on the observed pass-band, we treat these parameters in our fitting code as single values and fit for all transit mid-times ($T_{{\rm{C}}}$),  planet to star radius ratios ($R_{{\rm{p}}}/R_{\ast}$) and the coefficients responsible for the light curve systematics separately for each visit. In addition, we also fit $R_{{\rm{p}}}/R_{\ast}$ to a common value for the two G430L gratings. At each run, we used the derived three transit mid-times  and complemented them with all available transit times reported in the literature (with number of transit times indicated in parenthesis after the authors), namely the transit times reported in \cite{bakos07} (1), \cite{ johnson08} (7) and \cite{winn08} (2) and fitted a linear ephemeris of the form:

\begin{equation}
  T_{\mathrm{{{\rm{C}}}}}(E) = T_{\mathrm{0}} + E\times P,
\end{equation}       
where $T_{{{\rm{C}}}}$ is the central time at each observation, 
$T_{\mathrm{0}}$ is the reference transit time, 
and {\sl{E}} and {\sl{P}} are the transit epoch and 
period, respectively.

The result for the orbital period was then used as an initial guess for a new run. The complete process was iterated until a convergence for the value of the orbital period. Although we used the system parameters reported in \cite{johnson08} as initial guesses for the first run of our code, we found that {\tt{MPFIT}} is generally insensitive to the initial guess, except  the cases when the starting values significantly differ from the actual ones. 
 
We followed the aforementioned procedure and also investigated the results for all system parameters using the four limb darkening coefficients from the {\tt{1D}} and {\tt{3D}} atmospheric modelling. Generally both approaches provided results in good agreement (at the one sigma level) with a $\chi^2$ value slightly smaller ($<6\%$) for the derived parameters using the {\tt{3D}} limb darkening coefficients. Our best fitting model gave 
$a/r_{\ast}=9.853\pm0.071$, 
$i=85.634^{\circ}\pm0.056^{\circ}$, 
$\rho_{\ast}=0.908\pm 0.020~g\,cm^{-3}$. Table~\ref{tab:colordepthtab} summarises the
limb darkening coefficients used for each data set as well as the results for $R_{\rm{p}}/R_{\ast}$. The best fit models and detrended light curves for the three STIS and WFC3 white light curves are presented in Fig.~\ref{fig:wlc_stis} and Fig.~\ref{fig:wlc_wfc3}.  
We note that the systematics model identified individually for each white light curve differed from the ones identified from the joint fit of the four data sets. In the joint fit, the BIC was minimised for all STIS and WFC3 white light curves when we input models 4 and 3 from Table~\ref{tab:modseltab}, respectively. As the results for $i$, $a/R_{\ast}$ and   $R_{{\rm{p}}}/R_{\ast}$ were found to be within one sigma for each model, similar to those for the individual white light curves, as summarised in Table~\ref{tab:modseltab} we adopted models 4 and 2 and report the final white light curve fit result according to them.

For the transit ephemeris, we find the following results for the reference transit time and orbital period respectively:

\begin{equation}
 T_{\mathrm{0}} = 2453979.93202 \pm 0.00024\  {\rm{BJD}}
\end{equation}       

\begin{equation}
 P =  4.46529976\pm0.00000055\ {\rm{days}}.
\end{equation}       
The fit resulted in $\chi^{2}=13.36$ with 12 degrees of freedom (i.e. $\chi^{2}_r=1.11$), indicating a constant period. A plot of the observed minus computed transit times is displayed in Fig.~\ref{fig:ocfig}. We find no significant transit timing anomalies at a few minutes level. All transit times are reported in Table~\ref{tab:trtimestab} for convenience.

\begin{table}
\centering
\caption{Transit mid-times obtained from previous studies of \protect{\hat} along with the three measurements of $T_{C}$ acquired from the  {\it{HST}}/STIS data. The observed minus computed (O-C) residuals have been derived from a linear fit to the transit mid-times.}
\begin{tabular}{@{} r c c c }
\hline
\hline
Epoch & Central time & O$-$C & Refernce\\
 & (${\rm{BJD}}_{{\rm{TDB}}}$) & (days) & \\
\hline
    0 & $245   3979.93071\pm{ 0.00069}$ &  -0.001313 &        2\\
    1 & $245   3984.3978\pm{ 0.0090}$ &   0.000447 &        1\\
    2 & $245   3988.86274\pm{ 0.00076}$ &   0.000118 &        2\\
    4 & $245   3997.79277\pm{ 0.00054}$ &  -0.000452 &        2\\
    4 & $245   3997.79425\pm{ 0.00047}$ &   0.001028 &        2\\
    6 & $245   4006.72403\pm{ 0.00059}$ &   0.000208 &        2\\
    8 & $245   4015.6541\pm{ 0.0011}$ &  -0.000272 &        2\\
   20 & $245   4069.2387\pm{ 0.0029}$ &   0.000700 &        2\\
   86 & $245   4363.94677\pm{ 0.00091}$ &  -0.001031 &        3\\
   90 & $245   4381.8092\pm{ 0.0013}$ &   0.000249 &        3\\
  469 & $245   6074.15737\pm{ 0.00018}$ &  -0.000238 &        4\\
  470 & $245   6078.62307\pm{ 0.00018}$ &   0.000160 &        4\\
  478 & $245   6114.34537\pm{ 0.00020}$ &   0.000067 &        4\\
  495 & $245   6190.25542\pm{ 0.00018}$ &   0.000019 &        4\\
\hline
\multicolumn{4}{l}{Reference: $1-$\cite{bakos07}; $2-$\cite{winn07}; }\\
\multicolumn{4}{l}{$3-$\cite{johnson08}, $4-$this work. }
\label{tab:trtimestab}
\end{tabular}
\end{table}
We find overall agreement between the derived system parameters of \hat~in this work and the most recent results available in the literature. \cite{torres10} performed an uniform analysis of all ground based transit light curves coupled with the observable spectroscopic properties of \hat~and refined its physical parameters. For the main quantities that originate from a pure light curve fit: orbital inclination $(i)$, normalised semimajor axis $(a/R_{\ast})$, and stellar density $(\rho_{\ast})$ our results  differ with those of \cite{torres10} at the $\sim2\sigma$ level. Notably the absolute values for all of the three parameters are below the values reported in \cite{torres10}. Our measurements however, have much smaller uncertainties as they originate from a joint analysis of four transit observations, covering a complete transit, wide wavelength span and much higher precision compared to any of the present ground based photometric data sets and improve the precision of those quantities by a factor of a few. Finally, we employed limb darkening coefficients derived from a {\tt{3D}} stellar atmospheric models that explicitly take into account convective motions in the stellar surface granulation. That effect is neglected in the computation of limb darkening coefficients in the traditional {\tt{1D}} atmospheric models adopted in the previous analyses of \hat~data. 

When coupled with radial velocity data, transit photometry can provide a list of astrophysical parameters for a given planetary system. We therefore also complemented the STIS and WFC3 white light curves of \hat~with the existing radial velocity data measurements, reported in \cite{bakos07} and \cite{johnson08}. We excluded the radial velocity data that covers the Rossiter effect, as it is beyond the scope of this work to measure the spin orbit alignment of \hat. To perform a joint light curve and radial velocity modelling fit, we employed the publicly available {\tt{EXOFAST}} package of \cite{eastman12}, which incorporates a Markov Chain Monte Carlo analysis of the fitted parameters and uncertainties. {\tt{EXOFAST}} also served as an independent test for the propriety of our results and especially of the related uncertainties based on the {\tt{MPFIT}} procedure. The physical parameters and the improved ephemeris derived from the joint fitting with {\tt{MPFIT}} served as priors for {\tt{EXOFAST}}. This assumption is necessary because of a limitation in {\tt{EXOFAST}} that requires one to input a single band light curve in addition to the radial velocity data. Furthermore, we also modified the code to take into account the four parameter non linear limb darkening law instead of the currently available two-parameter quadratic limb darkening law. In addition, {\tt{EXOFAST}} utilises the \cite{torres10} empirical polynomial relation between the masses and radii of stars, and their surface gravity ($\log g$), effective temperatures ($T_{\rm{eff}}$) and metallicities [Fe/H], based on a large sample of non-interacting binaries with accurately measured astrophysical parameters. The empirical relation hence allows a derivation of a complete list of orbital and physical properties based on a joint analysis of transit photometric and velocity data \citep{eastman12}. We modelled each of the blue, red and near-infrared white light curves independently, except both STIS G430L curves which we fit simultaneously, as they originate from one band. We included the same models for the systematics as in the {\tt{MPFIT}} analysis. The final results of all parameters represent the average values from the three fits, except those for $i$, $a/R_{\ast}$ and   $R_{{\rm{p}}}/R_{\ast}$, which were computed using a weighted mean (as the {\tt{EXOFAST}} values were found in good agreement with the {\tt{MPFIT}} result and because we used four multi-epoch data sets). The transit ephemeris was derived using the approach described at the beginning of this section. Table~\ref{tab:rvrestab} summarises our final results. 

\begin{table}
\caption{System parameters for \protect{\hat}b and its hosts star based on the STIS, WFC3 and radial velocity data.}
\centering
\begin{tabular}{c@{\hskip 0.025in} c@{\hskip 0.025in}  c@{\hskip 0.025in}  }
\hline\hline
Symbol & Parameter \& Units & Value\\
\hline
\multicolumn{3}{l}{$Stellar$ $parameters$} \\
\hline
$M_{*}$ &Mass (\msun)  & $1.151_{-0.051}^{+0.052}$\\
$R_{*}$ &Radius (\rsun)  & $1.174_{-0.027}^{+0.026}$\\
$L_{*}$ &Luminosity (\lsun)  & $1.585_{-0.094}^{+0.099}$\\
$\rho_*$ &Density (cgs)  & $0.908_{-0.022}^{+0.019}$\\
$\log(g_*)$ &Surface gravity (cgs)   & $4.359_{-0.014}^{+0.014}$\\
$\teff$ &Effective temperature (K)  & $5980_{-49}^{+49}$\\
$\feh$ &Metalicity   & $0.130_{-0.008}^{+0.008}$\\
\hline
\multicolumn{3}{l}{$Planetary$ $parameters$} \\
\hline
$P$ &Period (days)  & $4.46529976{\pm(55)}$\\
$a$ &Semi-major axis (AU)   & $0.05561_{-0.00083}^{+0.00082}$\\
$M_{P}$ &Mass (\mj)   & $0.525_{-0.019}^{+0.019}$\\
$R_{P}$ &Radius (\rj)   & $1.319_{-0.019}^{+0.019}$\\
$\rho_{P}$ &Density (cgs)  & $0.282_{-0.009}^{+0.010}$\\
$\log(g_{P})$ &Surface gravity   & $2.873_{-0.010}^{+0.010}$\\
$T_{eq}$ &Equilibrium temperature ($K$)  & $1322_{-15}^{+14}$\\
$\Theta$ &Safronov Number  & $0.0384_{-0.0012}^{+0.0012}$\\
$\fave$ &Incident flux (\fluxcgs)   & $0.699_{-0.032}^{+0.032}$\\
\hline
\multicolumn{3}{l}{$RV$ $parameters$} \\
\hline
$K$  &RV semi-amplitude ($m\,s^{-1}$)   & $58.9_{-1.2}^{+1.2}$\\
$M_P\sin i$  &Minimum mass (\mj)   & $0.524_{-0.019}^{+0.019}$\\
$M_{P}/M_{*}$  &Mass ratio   & $0.000436_{-0.000011}^{+0.000011}$\\
\hline
\multicolumn{3}{l}{$Transit$ $parameters$} \\
\hline
$T_C$  &Transit time (\bjdtdb)   & $2453979.93202{\pm(24)}$\\
$R_{P}/R_{*}$  &Radius of planet in stellar radii   & $0.11802_{-0.00018}^{+0.00018}$\\
$a/R_{*}$  &Semi-major axis in stellar radii   & $9.853_{-0.071}^{+0.071}$\\
$i$  &Inclination (degrees)   & $85.634_{-0.056}^{+0.056}$\\
$b$  &Impact Parameter   & $0.7501_{-0.0069}^{+0.0064}$\\
$\tau$  &Ingress/egress duration (days)   & $0.02324_{-0.00047}^{+0.00047}$\\
$T_{14}$  &Total duration (days)  & $0.11875_{-0.00053}^{+0.00049}$\\
\hline
\label{tab:rvrestab}
\end{tabular}
\end{table}

\subsection{Transmission spectra fits}

The primary science goal of this project was to construct a low resolution optical to near-infrared transmission spectrum of \hat b and to pursue the prediction of strong optical absorbers such as sodium (observed through the (Na\,I doublet at $\lambda=5893$\,\AA), potassium and water vapour or alternatively optical high altitude atmospheric hazes, gradually enhancing the absorption as a function of decreasing wavelength (i.e. $\lambda < 6000$\,\AA) in the planetary atmosphere. To construct the complete transmission spectrum, we extract light curves from spectral bins from each \stis grating\footnote{we refer the reader to \cite{wakeford13} for details on the derivation of the WFC3 transmission spectrum.} on an individual basis, which enables an independent check for consistency. To perform the light curve fitting we rely on the {\tt{MPFIT}} procedure and  the system parameters derived from the joint analysis of the four white light curves  in Section~\ref{sec:syspar}. In particular, for each light curve fit, we fix the orbital period ($P$), inclination ($i$) and normalised semi-major axis ($a/R_{\ast}$) to their joint values along with the transit mid-times ($T_{\rm{C}}$) in each corresponding grating. Furthermore, we keep the four limb darkening coefficients fixed to their theoretical values. To construct the transmission spectrum in each grating, we fit for $R_{\rm{p}}/R_{\ast}$ and the parameters describing the instrument systematics in the baseline function. 

Model selection was performed for each grating assuming photometric errors based on pure photon noise. Once a given systematics model was selected, we re-estimated common uncertainties for our photometric data points, obtained from each spectral bin (i.e. taking the variance of the light curve residual). This is reasonable, because all photometric errors are equal in a given spectral bin as is typically the case for space-based observations. For example, we find no evidence in our data for a significant background variation which potentially could cause a variation of the photometric errors.

\begin{figure}
\includegraphics[trim = 0 0 0 0, clip, width = 0.48\textwidth]{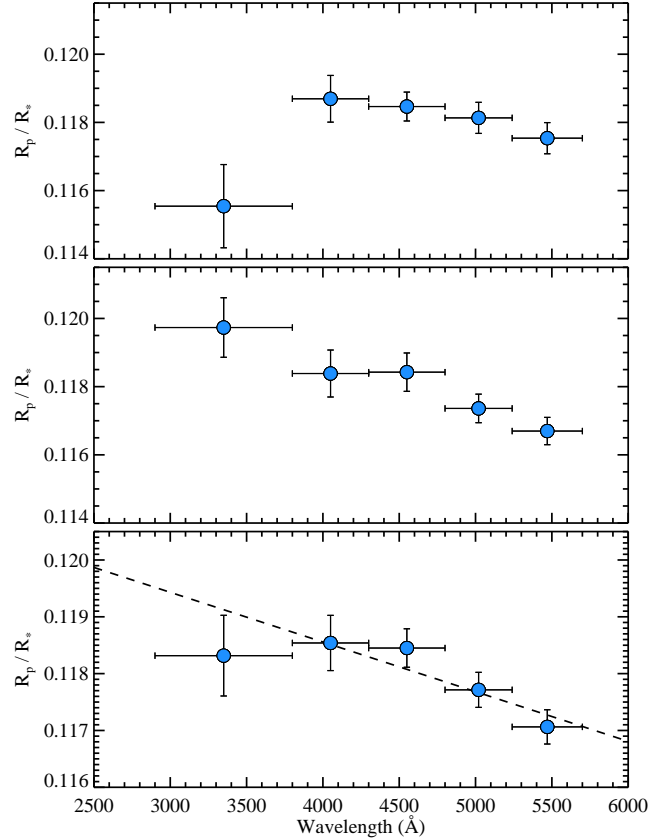}
\caption{{\it{Top panel:}} STIS G430L transmission spectra based on individual light curve fits to the data from Visit 7; {\it{Middle panel:}} Visit 8; {\it{Lower panel:}} weighted mean of Visit\,7 and 8, compared to a linear fit. The wavelength bin sizes are indicated with horizontal error bars and the 1$\sigma$ uncertainties for the measured planet to star radius ratio $(R_{pl}/R_{\ast})$ are indicated with vertical error bars.}
\label{fig:bluetranspecfig}
\end{figure}

\begin{figure}
\includegraphics[trim = 0 0 0 0, clip, width = 0.48\textwidth]{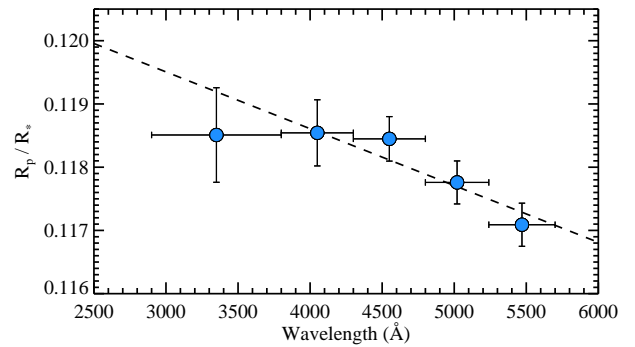}
\caption{Similar to  \protect{Fig.~\ref{fig:bluetranspecfig}}, but for the joint fit to visits\,7\,\&\,8.}
\label{fig:simbluetranspecfig}
\end{figure}
\subsubsection{G430L}
To construct the G430L transmission spectrum we produced light curves from spectral bins with custom sizes, common for both visits (7 and 8). Initially, we choose five bins with fixed size (in this case $\sim$ 500 \AA\,wide), as in various similar spectrophotometric studies. Although it seems natural to select an integer number and fixed size for the bins rather than various bin sizes, it has been found in the analysis that the light curves at that bin choice exhibited significant differences in the quality, except in the last two reddest bins. This is not surprising as the signal-to-noise of the data points in each light curve is correlated with the flux level within the wavelength range, constrained from the borders of each bin in the stellar spectrum and the sensitivity of \stis (see Fig.~\ref{fig:stis_spectrum}). We therefore choose to construct light curves from custom size bins, yet placing uniform constrain on the signal-to-noise ratio of the photometric measurements, as determined naturally by the spectra. Ideally one would aim to select a bin size such as not to lose too much information of the subsequent transmission spectrum in larger bins, nor to obtain light curves from tiny bins dominated by photon noise. 
We therefore experimented with various bin sizes, such that the signal-to-noise ratio of each lightcurve was $\sim(2-3)\times10^{3}$. Following this approach for both G430L data sets, we constructed five bin sizes, as summarised in  Table~\ref{tab:transpectab}. 

We measured the planet to star radius ratio ($R_{\rm{p}}/R_{\ast}$) with a fit to the transit light curves originating from each wavelength bin. It has been found during the analysis  that light curve fits with {\tt{3D}} limb darkening coefficients generally result in lower $\chi_{r}^2$ compared to the {\tt{1D}} alternative (typically less than $1-2\%$). We therefore report the remaining of the analysis based on the {\tt{3D}} limb darkening coefficients. Model selection was performed in two stages: i) first we used the data from each grating individually and ii) combined in a simultaneous fit. In the first case it has been found that the BIC values for all of the competing models were similar without clearly indicated models. The lowest BIC values were obtained when using models 4 and 1 (see Table~\ref{tab:modseltab} for details) for visits 7 and 8, respectively. However, it should be pointed out that both transmission spectra originating from the aforementioned models were remarkably similar in shape, i.e. the individual measurements agree within the $1\sigma$ error bars. Although both related data sets were obtained at about four months difference (see Table~\ref{tab:obsdatetab}), there is no evidence for a difference in the measured average level nor the shape of the transmission spectrum (in terms of $R_{\rm{p}}/R_{\ast}$). This is unsurprisingly given the fact that the host star is not found to be active and there is no evidence for spot crossing events, which might cause wavelength dependent transit-depth shifts. A significant exception however, resulted for the measured $R_{\rm{p}}/R_{\ast}$ based on the bluest bin ($2900-3800$ \AA), where the difference is at $\sim2\sigma$ level (see  Fig.~\ref{fig:bluetranspecfig} for details). Notably, it was only that measurement that varied with $\sim2\sigma$ during the model selection stage for visit 7 from its initial value in agreement with the value from visit 8 down to its lowest position (Models 1 to 4, respectively). In contrast, the same spectral bin resulted in similar $R_{\rm{p}}/R_{\ast}$ for visit 8 regardless {\rm{of}} the applied systematics model. While we could interpret the described behaviour of the visit 7 data set in the bin range $2900-3800$ \AA~as indicative for unreliable data, we chose not to ignore it and proceeded to a simultaneous fit of both visits as the light curves complement each other in phase when taken together.

\begin{figure*}
\includegraphics[width=\textwidth,height=\textheight,keepaspectratio]{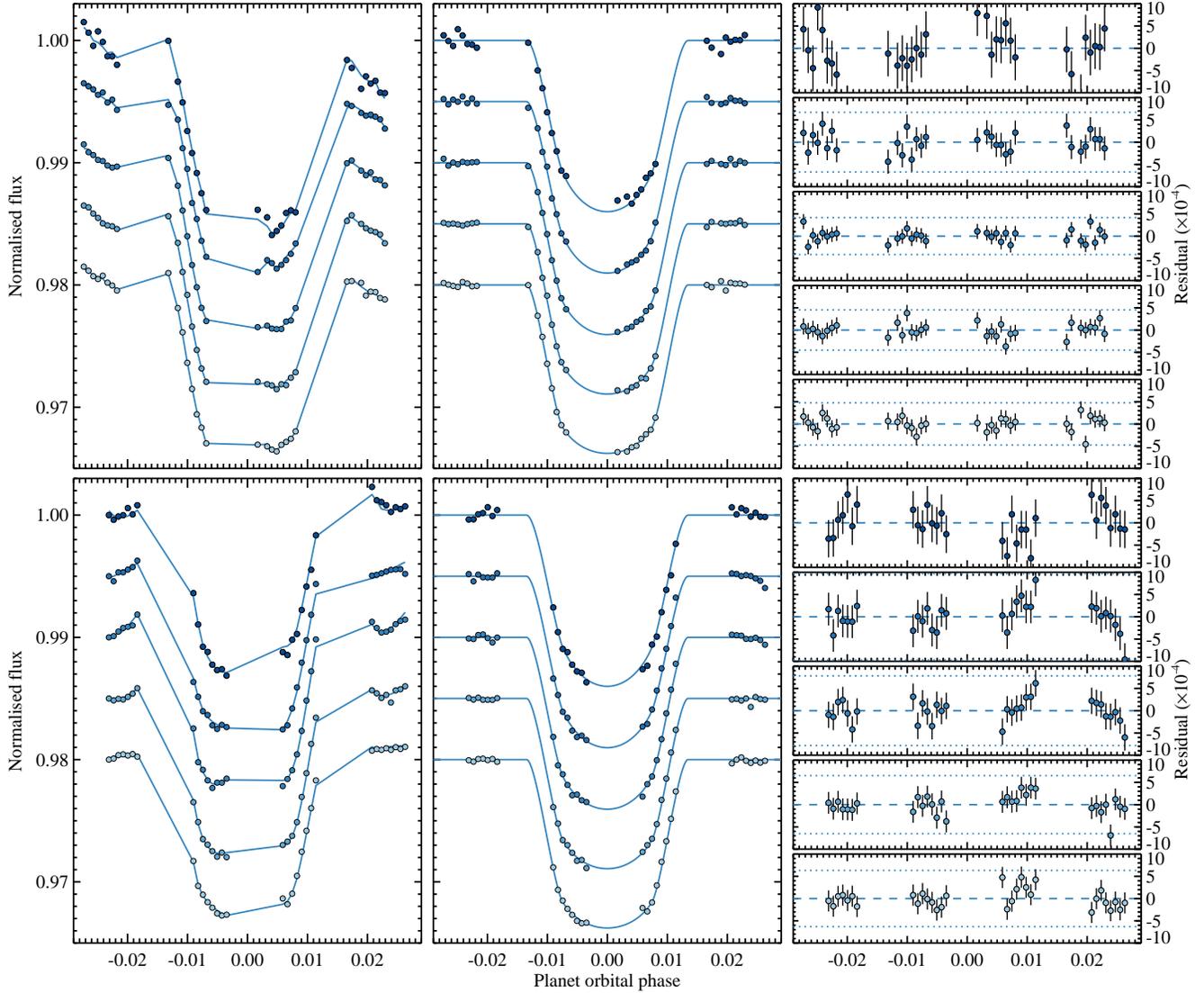}
\caption{\protect{{\hst}/{\stis}}~G430L observations on UT 2012 May 26 (Visit 7) and September 19 (Visit 8), top and lower three panels, respectively; {\sl{Left panels:}} raw light curves and the best-fit transit radius, multiplied in flux by a systematics model for each spectral bin, shifted with an arbitrary constant for clarity; {\sl{Middle panels:}} Corrected light curves and best-fit transit model; {\sl{Right panels:}} Observed minus computed residuals with error bars compared to a null level. A $3\sigma$ level is indicated with dotted lines.}
\label{fig:bluelcfig}
\end{figure*}

A simultaneous fit was performed to both blue visits aiming to derive a single value for the planet to star radius ratio ($R_{\rm{p}}/R_{\ast}$). We examined again the BIC evolution using the four mentioned systematics models and also included a fifth case, based on the best result from the individual model selection approach. Generally, the transmission spectrum did not change from model-to-model. Again $R_{\rm{p}}/R_{\ast}$  measured in the first bin was found to be the most unstable. However, in contrast to the first case, $R_{\rm{p}}/R_{\ast}$ remained within $\sim1\sigma$. The lowest BIC value was obtained for case model 5, i.e. models 4 and 1 for visits 7 and 8, respectively. In fact, this result is a confirmation to our previous finding in the individual model selection method. We therefore chose to complete the analysis of the blue part of the STIS transmission spectrum using those models.  Finally, and to corroborate our results for the G430L transmission spectrum, we also shifted the spectral bins around their present position with 50 \AA\,and examined the resulting transmission spectrum, finding no significant differences (i.e. larger than $1 \sigma$). Table~\ref{tab:transpectab} and   Fig.~\ref{fig:simbluetranspecfig} report our final results. We further display the raw and detrended light curves along with the residuals in Fig.~\ref{fig:bluelcfig}.

\begin{table*}
\centering
\caption{Measured $R_{\rm{p}}/R_{\ast}$ from transit light curves fits and theoretical limb darkening coefficients from {\tt{3D}} stellar atmosphere models.}
\begin{tabular}{@{} c c c c c c}
\hline
\hline
$\lambda$ (\AA)   &   $R_{\rm{p}}/R_{\ast}$     &  $c_{1}$  &   $c_{2}$     &    $c_{3}$   &   $c_{4}$  \\

\hline
 Visit 7 \& 8 & & & & & \\
$2900 - 3800$ & $0.11851 \pm 0.00075$ & 0.2344 & 0.6735 & 0.0508 & -0.0574\\
$3800 - 4300$ & $0.11854 \pm 0.00052$ & 0.3343 & 0.4421 & 0.2378 & -0.1260\\
$4300 - 4800$ & $0.11845 \pm 0.00035$ & 0.4306 & 0.4531 & 0.0111 & -0.0289\\
$4800 - 5240$ & $0.11776 \pm 0.00034$ & 0.5291 & 0.2790 & 0.0639 & -0.0424\\
$5240 - 5700$ & $0.11709 \pm 0.00034$ & 0.6016 & 0.1090 & 0.1561 & -0.0686\\
Visit 20 & & & & & \\
$5293 - 5878$ & $0.11826 \pm 0.00039$ & 0.6153 & 0.0837 & 0.1617 & -0.0700\\
$5878 - 5908$ & $0.1223 \pm 0.0014$ & 0.6309 & 0.0098 & 0.2242 & -0.0975\\
$5908 - 6493$ & $0.11778 \pm 0.00036$ & 0.6705 & -0.0572 & 0.2281 & -0.0913\\
%
$6500 - 6900$ & $0.11782 \pm 0.00049$ & 0.7170 & -0.1751 & 0.2844 & -0.1106\\
$6900 - 7300$ & $0.11799 \pm 0.00053$ & 0.7246 & -0.2432 & 0.3483 & -0.1326\\
$7300 - 7800$ & $0.11825 \pm 0.00049$ & 0.7374 & -0.3021 & 0.3787 & -0.1421\\
$7800 - 8500$ & $0.11817 \pm 0.00050$ & 0.7507 & -0.3606 & 0.4072 & -0.1512\\
$8500 - 9200$ & $0.11868 \pm 0.00071$ & 0.7502 & -0.4111 & 0.4339 & -0.1594\\
$9200 - 10200$ & $0.11988 \pm 0.00046$ & 0.7472 & -0.4208 & 0.4174 & -0.1494\\
\hline
\label{tab:transpectab} 
\end{tabular}
\end{table*}

\begin{figure*}
\includegraphics[width=\textwidth,height=\textheight,keepaspectratio]{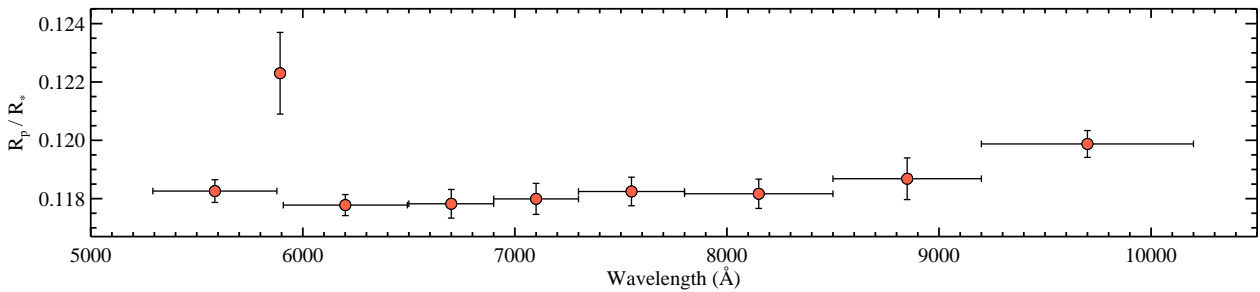}
\caption{ STIS G750L transmission spectrum of \protect{\hat}b.}
\label{fig:redtranspecfig}
\end{figure*}

\begin{figure*}
\includegraphics[width=\textwidth,height=\textheight,keepaspectratio]{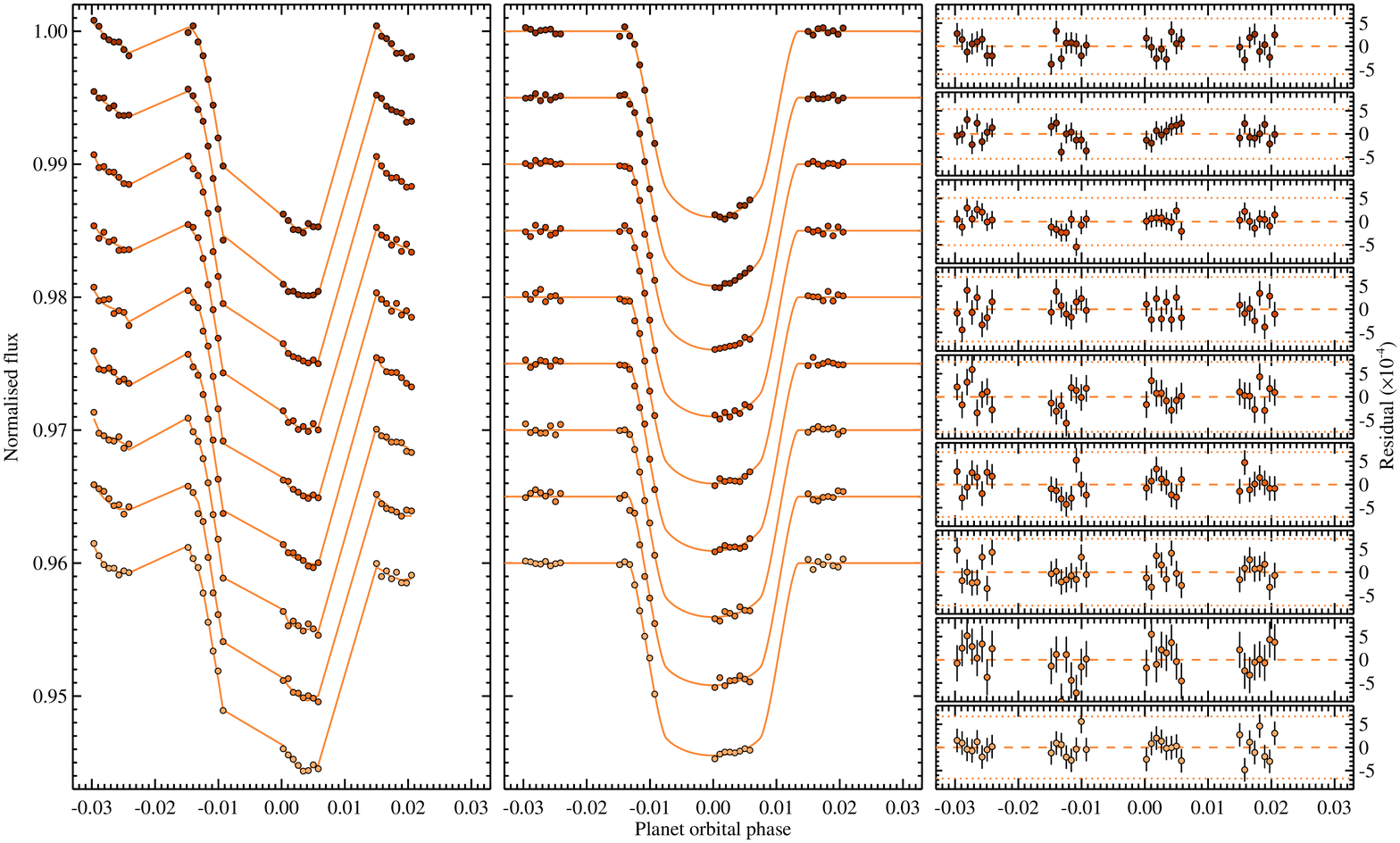}
\caption{Same as Fig.~\ref{fig:bluelcfig}, but for the {\it{HST}}/STIS G750L data, obtained on UT 2012 May 30 (Visit 20).}
\label{fig:redlcfig}
\end{figure*}

\subsubsection{G750L}\label{sec:g750sec}
The red STIS data were analysed similar to the blue. Again, we choose custom bin sizes through the complete wavelength region yet aiming to obtain similar signal-to-noise ratios for each light curve in the range $(2-3)\times10^3$. This approach ensures that each spectral bin produces light curve that is similar in quality to the remaining light curves from the entire grating. 

Model selection was performed similar to grating G430L, attempting the four analytical models for systematics (see Table~\ref{tab:modseltab} for details). At each trial we constructed the transmission spectrum of \hat b in terms of  $R_{\rm{p}}/R_{\ast}$. Overall, the spectrum remained quite similar in shape and each measurement varied within $1\sigma$ throughout the model selecting process. The BIC statistic was minimised for Model 2. We therefore performed the detrending of each bin of the G750L grating with that model. Similar to the analyses of the transmission spectrum of G430L, we also experimented with shifted spectral bins ($\pm50$ \AA) to secure stability of the derived spectrum and found good agreement between the $R_{\rm{p}}/R_{\ast}$ measurements from the three trials. Comprehensively we did not find a significant difference between $\chi^2$ values originating from the fits based on the {\tt{1D}} against {\tt{3D}} limb darkening coefficients due to the longer wavelengths covered by the G750L grating, where the variation of the limb darkening with wavelength is noticeably weaker. However, for clarity and consistency we report the results using the {\tt{3D}}  models. 

\stis fringe corrected versus uncorrected time series were employed to investigate the impact of the fringe effect on the resulting transmission spectra. Similar to \cite{knutson07} we found that the fringe effect do not influence significantly the transmission spectrum when using large bin sizes (such as our three reddest bins). We choose to report the G750L transmission spectrum based from the fringe corrected data for clarity and consistency.

The transmission spectrum is marginally flat with increasing absorption longward of $\sim8500$ \AA~and a deviation of the bin that brackets the sodium feature. As we describe in detail this part of the analysis in Section~\ref{sec:sodiumsec}, we emphasise that we report the wavelength range $\lambda = 5290-6500$~\AA~of the spectrum with the results from the sodium search. The final version of the G750L spectrum is summarised in Table~\ref{tab:transpectab} and  Fig.~\ref{fig:redtranspecfig}. The raw and corrected light curves used to construct the transmission spectrum along with the residuals are exhibited in Fig.~\ref{fig:redlcfig}.

\subsubsection{Sodium}\label{sec:sodiumsec}
The most readily identifiable feature in the optical transmission spectrum of \hat b is the sodium Na\,I doublet. We pursued the sodium signature employing the differential light curve procedures described in \cite{charbonneau02, sing08b, huitson12}. In summary, we first produced raw photometric time series summing the detected counts over spectral bins, which were then corrected for systematics, that depend on the $HST$ orbital phase (fourth order plinomial)
and a linear time term ($t$). The detection of the individual lines of the sodium resonance doublet ($\lambda_{{\rm{Na\,I}}} = 5890$~\AA~and $\lambda_{{\rm{Na\,II}}} = 5896$~\AA), separated only by $\sim6$ \AA~is hampered by the small scale of the \stis instrument  ($\sim 5$ \AA~pixel$^{-1}$) at low spectral resolution ($R=500$). In addition, the precise width of the feature we seek is unknown. We therefore, select bands of varying width, each centred on the sodium doublet ($\lambda_{{\rm{Na}}} = 5893$~\AA), aiming similar to \cite{charbonneau02} to span ``narrow", ``medium" and ``wide" wavelength regions. However, instead of using three bands we choose to perform a more detailed investigation of the evolution of the detection/non-detection of the sodium feature. The set of bands centred on the sodium feature are 15, 30, 45, 60, 75 and 90~\AA~wide.  We also coupled each of these bands with ``blue" and ``red" bands with a width of $600$~\AA, in order to bracket the ``center" band. We selected the out-bands to be significantly wider compared to the in-bands to secure stability of these reference measurements. We further produced photometric time series for each of these eighteen bands as described in Section~\ref{sec:g750sec}. To investigate for the sodium feature we first computed the mean light curve of the blue ``b(t)" and red ``r(t)" bands and subtracted that light curve from the light curve of the centre ``c(t)" band: $D_{{\rm{Na}}}(t) = c(t) - [b(t)+r(t)]/2$, where $D_{{\rm{Na}}}(t) $ is the differential light curve. As pointed out by \cite{charbonneau02}, the advantage of this linear combination is that it removes most of the variations due to the colour dependence of the limb darkening of the stellar continuum.  We then computed the difference in the mean of the in- and out-of-transit data (indicated by the relevant subscripts):  $\Delta D_{Na} = \overline{ D_{Na}(t_{in}) } - \overline{D_{Na}(t_{out}) } $. To estimate an uncertainty of the $\Delta D_{Na} $ differences, we computed the standard deviation of the mean, \cite{bevington03}. Following this procedure, we found that the sodium feature is detected in each of the six bands with a significance ranging as low as $1.2\sigma$ for the 45~\AA~band to as high as $3.3\sigma$ for the 30~\AA~band. We therefore report the exact values for the highest detection significance:  $\Delta D_{Na}=(-98\pm30)\times10^{-5}$. Furthermore, for the six central bands, starting with the narrowest to the widest we observe a monotonic decrease of the measured standard deviation of the out-of-transit data. These values are close to the predictions of photon noise-limited precision. 

 We also used the aforementioned procedure with six bands and scanned the wavelength region $+/-50$~\AA~from the sodium resonance line in steps of $5$~\AA, which is close to the resolution of the \stis instrument. Sodium was detected only in the $30$~\AA~region, centred on $\lambda_{{\rm{Na}}}=5893$~\AA~and most significantly (i.e. strongest signal) when using the narrow bands. In addition,  no signal was detected when we moved the bands more than $+/-30$~\AA~away from the sodium resonance line.

\begin{figure*}
        \centering
 
        \begin{subfigure}
                \centering
                \includegraphics[scale=.5]{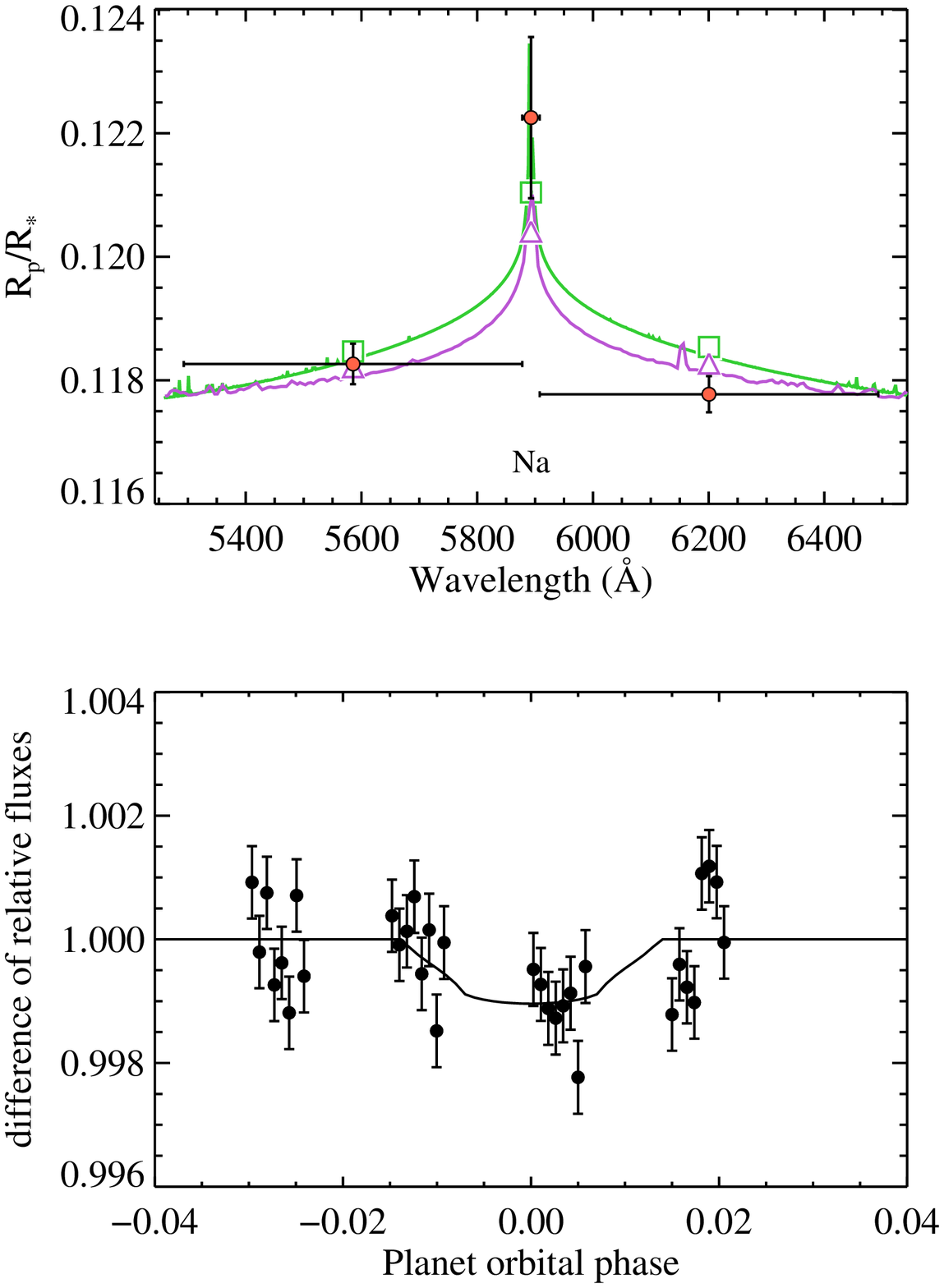}
        \end{subfigure}
         \begin{subfigure}
                \centering
                \includegraphics[scale=.5]{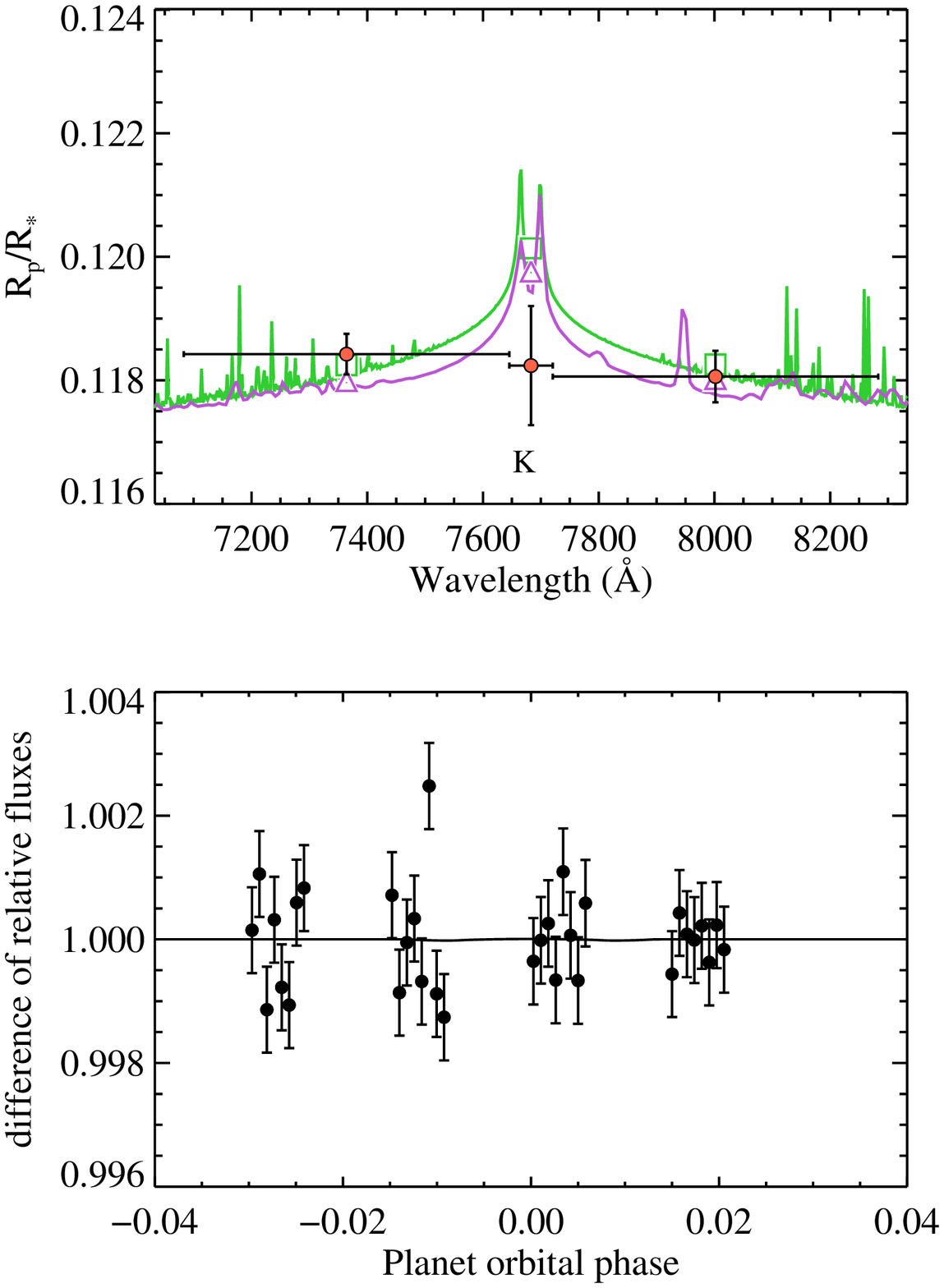}
        \end{subfigure}
 
         \caption{ {\it{Top panels:}} The observational transmission spectrum (in terms of $R_{\rm{p}}/R_{\ast}$ obtained from light curve fits, shown with dots and vertical error bars; horizontal errors indicate the band widths used to obtain light curves)  around the sodium and potassium features (left and right panels, respectively), compared to theoretical cloud free models (shown with green and purple continuous lines for \protect\cite{fortney08,fortney10} and \protect\cite{burrows10} and \protect\cite{howe12}, respectively described in detail in Section~\ref{sec:discussionsec}) and the related radius predictions (boxes and triangles), both shifted from each other and the data for clarity; {\it{Lower panels:}} Differential-transit light curves of 30 and 75~\AA~bands, centred on the sodium and potassium doublets, respectively to that of a reference composed from the average of a blue and red band, bracketing each feature. Differential limb-darkening curves are indicated with continuous lines. A colour version is available in the online version of the journal.}
\label{fig:sodiumpotassiumfig}
\end{figure*}
Although, the differential procedure significantly decreases the contribution of the stellar limb darkening to the variation of the transit depth, we must also quantify how much of the observed decrement is a result of the distinctive limb darkening exhibited by the sodium line relative to the adjacent continuum, calculating the differential limb darkened transit light curves, As evident in Fig.~\ref{fig:sodiumpotassiumfig} (lower left panel) the modelled differential time series shows only weak residual limb darkening effects. 

In addition to the positive result from the differential method, we also detected the sodium feature by comparing the measured planet radii from individual fits of transit  and systematics models (as in Section~\ref{sec:g750sec}) to the six raw light curves from the blue, centre and red bands, respectively. This result is not surprising, given the accurate prediction of the stellar limb darkening, based on {\tt{3D}} atmospheric models. Fig.~\ref{fig:sodiumpotassiumfig} (left panels) summarises the sodium detection from both approaches described above. 

Based on the result for a significantly deeper transit in the sodium band ($3.3\sigma$) compared to the adjacent red and blue bands and the negligible contribution of the residual limb darkening, we conclude that we have detected sodium in the planetary atmosphere of \hat b.

\subsubsection{Potassium and H$\alpha$}\label{sec:potassiumsec}
We followed the two methods, described in Section~\ref{sec:sodiumsec} to inspect the \stis data for potassium. Because the potassium spectral signature is known to be formed by a doublet, centred at {\rm{$\lambda_{K} \approx 7684$}} \AA~with two cores separated by 34~\AA, we first employed only the set of medium and wide bands, i.e. 45, 60, 75 and 90~\AA, centred at $\lambda_{K} $. Each of the four bands showed no detection of transit depth decrement due to potassium when applying both methods. We therefore report the result for the 75~\AA~wide band that brackets the doublet: $\Delta D_{K} =(1\pm28)\times10^{-5}$ (see the right-hand panels in Fig.~\ref{fig:sodiumpotassiumfig}). The out-of-transit data resulted in standard deviation of $6.4\times 10^{-4}$ which is close to the expectation of photon noise-limited precision. 
We further investigated each of the potassium cores individually with the complete set of bands, i.e. including the two narrow ones. In the case of {\rm{$\lambda_{K\,D2} = 7667$}} \AA, we find no evidence of a significant variation in the transit depth, with the highest value at the 60~\AA~wide band with $\Delta D_{K\,D2}=(-25\pm24)\times10^{-5}$ and limb darkening also consistent with no variation: $\Delta D_{ld} =(12\pm16)\times10^{-5}$. The second core {\rm{$\lambda_{K\,D1} = 7701$}} \AA~also gave results consistent with no variation, indicating no presence of potassium. Based on these results we therefore conclude that there is no evidence in our data for the presence of potassium in the atmosphere \hat b.   

We also inspected the red \stis data for extra absorption in H$\alpha$ line (6563 \AA), following the aforementioned approach. We found no variation of the transit depth at that wavelength. 

\subsubsection{Transmission spectra level differences}\label{sec:leveldiffsec}
\begin{figure*}
\includegraphics[scale=0.6]{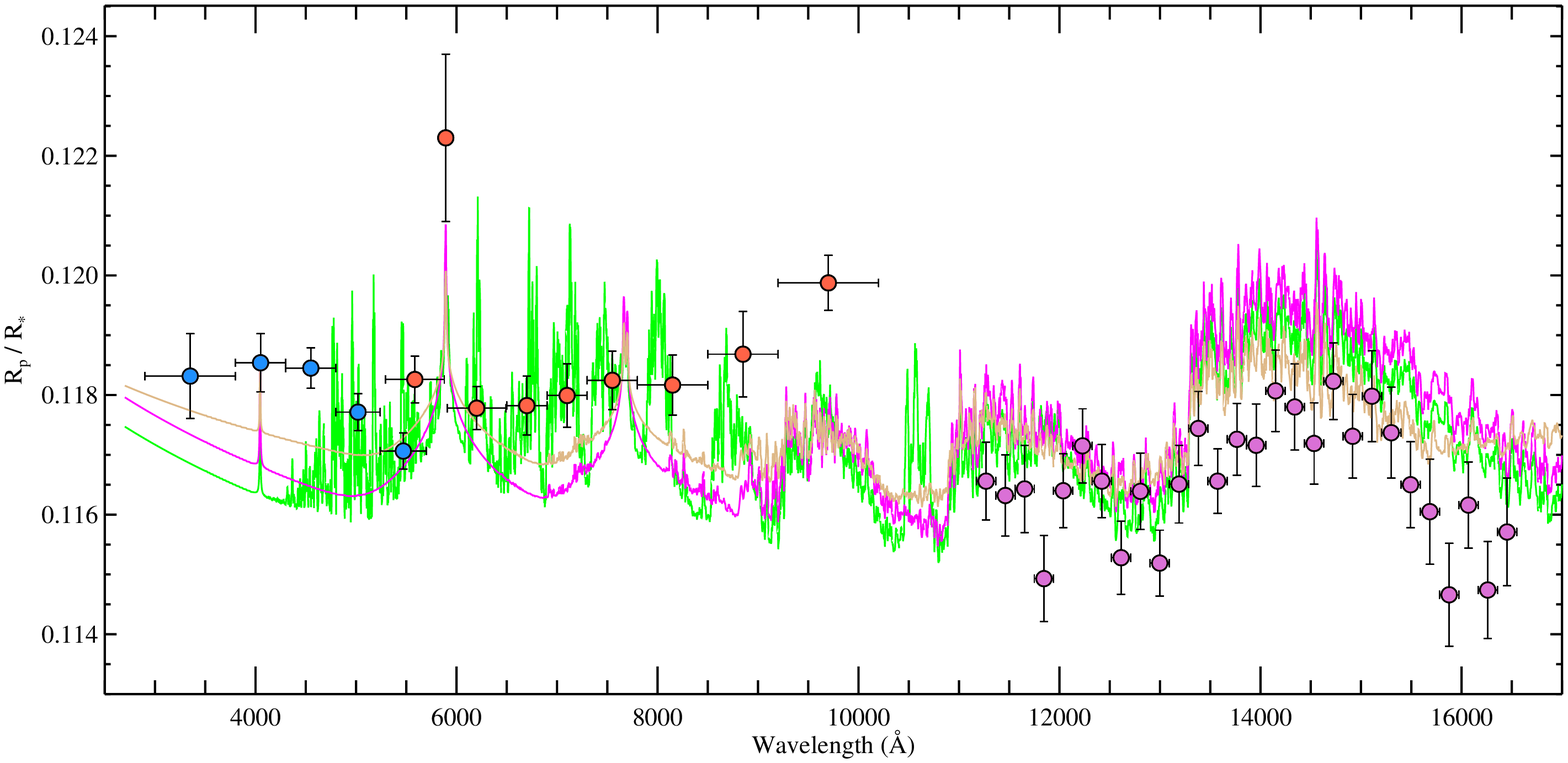}
\caption{{\it{HST}} transmission spectrum (blue, red and purple dots refer to the \stis G430L, G750L and \wfc data, respectively) compared to atmospheric models (continuos lines) generated for the \hat~system. \protect\cite{fortney08,fortney10} isothermal hydrostatic uniform abundance models with an equilibrium temperature 1000~K (brown line) and 1500~K (green line with TiO and VO; purple line: without TiO and VO). Horizontal and vertical error bars indicate the spectral bin sizes and the $R_{\rm{p}}/R_{\ast}$ uncertainties, respectively. A colour version is available in the online version of the journal.}
         \label{fig:isomod1500}
\end{figure*}

We complement the blue and red STIS transmission spectra discussed in this work with the recently reported near-infrared transmission spectrum of \hat b~by \cite{wakeford13}, which shows a significant absorption above the $5\sigma$ level, matching the 1.4~$\mu$m water absorption band. To secure a uniform analysis of the \stis and \wfc transmission spectra we coordinated our efforts with \cite{wakeford13} and employed the same values for the semimajor axis ($a/R_{\ast}$), orbital period ($P$) and inclination ($i$), as described in Section~\ref{sec:analysis}, in conjunction with limb darkening coefficients derived from the {\tt{3D}} stellar atmosphere models. Despite the uniform analysis, the level of the optical (blue and red) spectrum seem to be significantly higher than the near-infrared spectrum. Based on current giant planet atmospheric models, including \cite{burrows10} and \cite{howe12} or \cite{fortney08,fortney10}, described in detail in Section~\ref{sec:discussionsec}, the water feature is expected to peak higher than the base of the wings of the sodium and potassium resonance doublets (see Fig.~\ref{fig:daysidemod} and~\ref{fig:isomod1500}). Surprisingly, in the case of \hat b the water feature peaks nearly at the level of the wing bases of the sodium and potassium doublets (see $R_{\rm{p}}/R_{\ast}$ for systematic models 2 and 3 for the red \stis~and the \wfc~data, respectively in Table~\ref{tab:modseltab} in the appendix or Table~\ref{tab:colordepthtab}). Clearly, an alternative statement can be formulated presuming a higher optical level compared to a low water feature. Regardless the choice of formulation however, a fundamental question can be raised about the possible shifts in altitude between the three transmission spectra. One speculation for the observed difference could be attributed to stellar activity and spot crossing events. However, based on our stellar activity monitoring (see Section~\ref{sec:stellaractivity}) we see no evidence in our monitoring data (Fig.~\ref{fig:variability_plot} with Ca~II~H\&K chromospheric line emission index of $\log{R'_{H\&K}}=-4.984$, \cite{knutson10}) for variability of \hat~above $\pm 0.5\,\%$ (a scatter of 2 mmag consistent with the photometric noise), which is also compatible with the low chromospheric activity concluded in \cite{bakos07} and the fact that we do not observe any spot crossings during the transits. This hypothesis hence, cannot explain the observed difference of $4.3\pm1.6\,H$, where $H = 414\,km$ is the assumed scale height of \hat b. Another explanation for the observed difference could be a systematics in the data reduction stage. To help test this hypothesis and to place an absolute uncertainty on the levels of the transmission spectra we investigated the variation of the white light $R_{\rm{p}}/R_{\ast}$ measurements (including the known correlation of $R_{\rm{p}}/R_{\ast}$ with $a/R_{\ast}$ and $i$), as we varied $a/R_{\ast}$ and $i$ within their best-fit values with uncertainties (see Section~\ref{sec:syspar}). In particular, while keeping $a/R_{\ast}$ as a free parameter, we fixed the value of the orbital inclination $(i)$ to its lower and upper limits and obtained two radius estimates. The same approach gave two radius estimates when fixing $a/R_{\ast}$ and varying $i$. We then fixed both $a/R_{\ast}$ and $i$ to their upper and lower values, respectively and obtained another two radius estimates; the final two estimates were obtained while keeping one of the parameters to its low/high value while fixing the other to its high/low value. 

Using the eight  $R_{\rm{p}}/R_{\ast}$ estimates for each grating/grism, we further computed their mean and standard deviation values (see Table~\ref{tab:radvarctab}). Two interesting facts can be established after an inspection of Table~\ref{tab:radvarctab}. First, the \wfc data results in a significantly lower $R_{\rm{p}}/R_{\ast}$ value, compared to the \stis data. In fact, we also investigated the influence of the differential (with respect to \hat B companion star) versus single flux white light curve obtained from the \wfc data. Both results for $R_{\rm{p}}/R_{\ast}$ were found in good agreements (within their uncertaintes) with slightly higher value from the differential light curve; and second,  contrary to our results at the systematic model selection stage (see Table~\ref{tab:modseltab} in the appendix), the \wfc white light curve showed the smallest spread in  $R_{\rm{p}}/R_{\ast}$, followed by the red and blue \stis data. We estimate the differences $(\gamma)$ between the mean $R_{\rm{p}}/R_{\ast}$ levels for each of the three possible pairs of gratings/grisms of Table~\ref{tab:radvarctab} and report them in Table~\ref{tab:gamatab}. We estimated the related uncertainty by propagating the mean uncertainties (based on the eight $R_{\rm{p}}/R_{\ast}$ estimates) for each grating/grism. While both \stis data sets were found to be in agreement, our results implies a signifiant difference between the \stis and \wfc data at a $3\sigma$ level. 

\begin{table}
\centering
\caption{Mean and standard deviation values of $R_{\rm{p}}/R_{\ast}$ obtained while varying $a/R_{\ast}$ and $i$.}
\begin{tabular}{@{} l l l }
\hline
\hline
Data set   & $<R_{\rm{p}}/R_{\ast}>$     &  Deviation \\
\hline
Visit7/G430L   & 0.1192 & 0.0011\\
Visit20/G750L & 0.11826 & 0.00057\\
Visit8/G430L   & 0.1181 & 0.0011\\
Visit26/G141  & 0.11665 & 0.00020\\
\hline
\label{tab:radvarctab} 
\end{tabular}
\end{table}

\begin{table}
\centering
\caption{Transmission spectra level differences.}
\begin{tabular}{@{} c c }
\hline
\hline
Grating/grism pair   & $\gamma= R_{\rm{p1}}/R_{\ast} - R_{\rm{p2}}/R_{\ast}$\\
\hline
G430L -- G750L  & $0.00037\pm0.00079$\\
G430L -- G141    & $0.00197\pm0.00077$\\
G750L -- G141   &  $0.00161\pm0.00053$\\
\hline
\label{tab:gamatab} 
\end{tabular}
\end{table}

\begin{figure}
\includegraphics[trim = 0 0 0 0, clip, width = 0.48\textwidth]{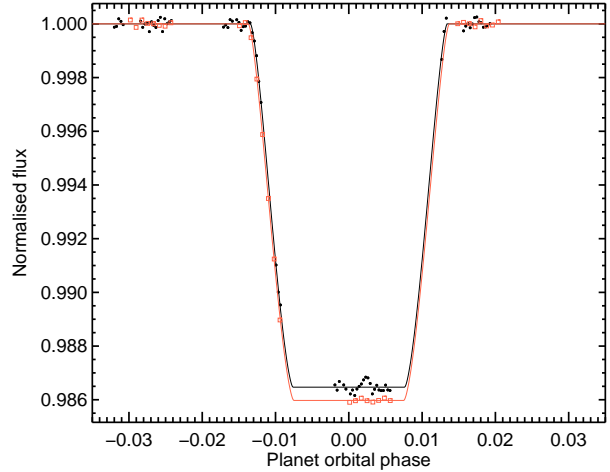}
\caption{An empirical demonstration of the transit depth difference between the red \stis (boxes) and \wfc (dots) limb darkening corrected white light curves and transit models (continuous lines). The difference is obvious in each transit exposure, including the ingress. A colour version is available in the online version of the journal.}
\label{fig:empericalshiftfig}
\end{figure}

\section{Discussion}\label{sec:discussionsec}
Placed together, the blue and red \stis transmission spectra exhibit several noticeable features. 
Based on the two methods, described in Section~\ref{sec:sodiumsec} and~\ref{sec:potassiumsec}, we conclude that we detected sodium (at the $3.3\sigma$ level at $\lambda=5893$~\AA), but no excess absorption due to potassium nor Balmer H$\alpha$ (at $\lambda = 7683$~\AA~and $6563$~\AA, respectively) in the \hat b planet atmosphere. The blue spectrum exhibits a gradual increase toward shorter wavelengths, while the red spectrum is marginally flat with an enhanced absorption beyond $\sim8500$~\AA. In what follows, we discuss the astrophysical implications of the individual and combined \stis and \wfc transmission spectra on \hat b's planetary atmosphere.

\subsection{Comparisons to existing theoretical models}\label{sec:thmodels}
The complete optical to near-infrared transmission spectrum was compared to two different sets of atmospheric models, similar to the analysis of \cite{wakeford13} and \cite{huitson13}; One set based on the formalism of  \cite{burrows10} and \cite{howe12} and the other, based on the formalism of \cite{fortney08,fortney10}. Our choice to employ two independently derived sets of atmospheric models is governed by the fact that not all model sets in the literature agree, as they were computed using different methods (see \citealt{shabram11} for details). To perform a comparison between theory and observation we employed pre-computed models and proceeded as follows. From each model, we averaged the model within the wavelength bins that were used to construct the observational transmission spectrum of \hat b (as detailed in Table~\ref{tab:transpectab}). Then the pre-computed models were fit to the data with a single free parameter that controls their vertical position, i.e. without any influence on the model shape. For each model we computed the $\chi^2$ statistic to quantify model selection. As the models are pre-computed, the number of degrees of freedom at this stage of the analysis is constant for each model, i.e. $DoF = n-m$, where $n$ is the number of data points (see the following text for each sets of models) and $m$ is the number of fitted parameters, i.e. $m=1$. The comparisons help to test the underlying model assumption (e.g. the presence/absence of TiO) and identify atmospheric species in the transmission spectrum.

The models from \cite{burrows10} and \cite{howe12} were specifically generated for the \hat~system using a {\tt{1D}} dayside temperature~--~pressure \tp profile with stellar irradiation, in radiative, chemical, and hydrostatic equilibrium. Chemical mixing ratios and corresponding opacities assume solar metallicity and local thermodynamical chemical equilibrium accounting for condensation with no ionisation, using the opacity database from \cite{sharp07} and the equilibrium chemical abundances from \cite{burrows99} and \cite{burrows01}.

The models based on \cite{fortney08,fortney10} were also specifically generated for the \hat~system and included a self-consistent treatment of radiative transfer and chemical equilibrium of neutral and ionic species. Chemical mixing ratios and opacities assume solar metallicity and local chemical equilibrium, accounting for condensation and thermal ionisation though no photochemistry \citep{lodders99, lodders02a,lodders02b, lodders06, visscher06, freedman08, lodders09}. In addition to isothermal models, transmission spectra were calculated using {\tt{1D}} temperature-pressure \tp profiles for the dayside, as well as an overall cooler planetary-averaged profile. Models were also generated both with and without the inclusion of TiO and VO opacities. Because each of the aforementioned models started at wavelength $\lambda = 3500$~\AA~(due to a restriction of the employed opacity database), we extrapolated the models in the range $2700-3500$~\AA~to enable self-content comparisons to the bluest \stis measurements along with all models.

\begin{figure*}
\includegraphics[scale=0.6]{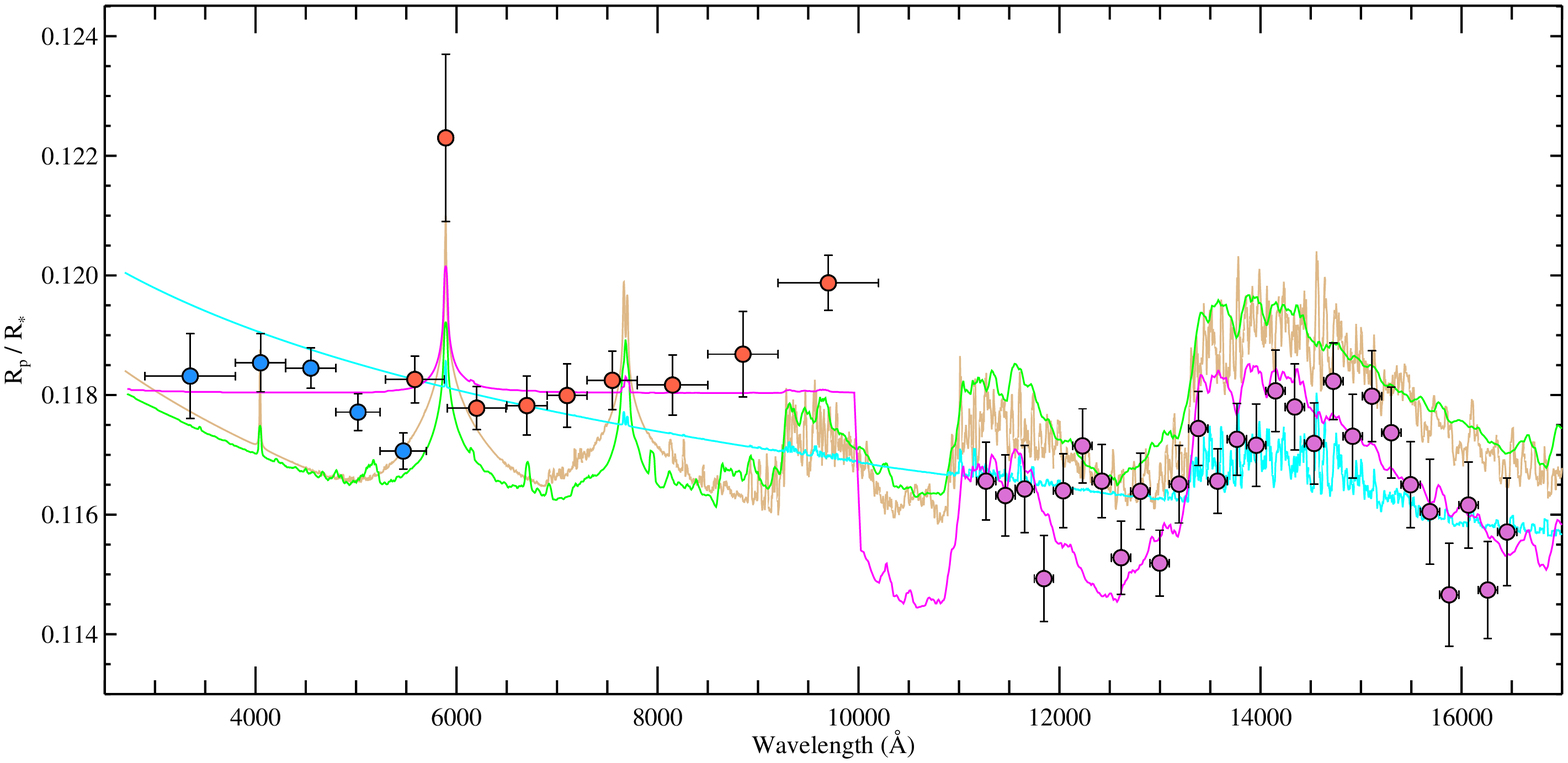}
         \caption{ Same as Fig.~\protect\ref{fig:isomod1500}, but for the models assuming specific pressure-temperature profiles and isothermal models with additional adjustments. Green line: dayside model from \protect\cite{burrows10} and \protect\cite{howe12} at solar metallicity without TiO/VO; Brown line: dayside model from \protect\cite{fortney08,fortney10} without TiO/VO. Purple line: an isothermal model of \protect\cite{burrows10} with an `extra absorber' at altitude with an opacity of $0.03~cm^2 g^{-1}$ from 0.4 to 1.0~$\micron$. Cyan line: isothermal model from \protect\cite{fortney08,fortney10} with enhanced (factor of $10^3$) Rayleigh scattering. Horizontal and vertical error bars indicate the spectral bin sizes and the $R_{\rm{p}}/R_{\ast}$ uncertainties, respectively.} 
         \label{fig:daysidemod}
\end{figure*}

To compare the standard theory with observation, we first employed \cite{burrows10} and \cite{howe12} dayside model in hydrostatic equilibrium at solar metallicity without TiO/VO and the the dayside averaged model of \cite{fortney08,fortney10}. Although both models were pre-computed for the system parameters of \hat, assuming cloudless atmosphere, they resulted in poor fits. The dayside averaged model of \cite{fortney08,fortney10} produced the lowest $\chi^2 = 231$, followed by the model of \cite{burrows10} and \cite{howe12} $\chi^2 = 311$ with 41 degrees of freedom (see\,Fig.~\ref{fig:daysidemod}).

A comparison of the observed transmission spectrum was also carried out to isothermal hydrostatic uniform abundance models of \cite{fortney08,fortney10} with two equilibrium temperature regimes, different from the one of \hat b (i.e. $\sim1200\,K$).  Such analysis helps provide a general understanding of the observed features as well as any departures from them. The first regime included isothermal models with equilibrium temperature of $T_{eq}=1500\,K$, (to represent the hotter dayside) with and without TiO/VO. The second regime included a model at lower equilibrium temperature $T_{eq}=1000\,K$ that represents a cooler terminator and not containing TiO/VO. The fit to the $T_{eq}=1000\,K$ isothermal model is the best from the three isothermal models fit to the data, with $\chi^2 = 168$ and 41 degrees of freedom (see Fig.~\ref{fig:isomod1500}). In fact, a temperature of $1000$~K is also consistent with the result from \cite{wakeford13}. The TiO/VO containing model gave a lower $\chi^2$ than the non-TiO/VO model ($\chi^2= 218$ and 324, respectively), because of the enhanced absorption of the TiO/VO model in the range $\lambda \sim 4000-9000$~\AA~(see Fig.~\ref{fig:isomod1500}). The presence of TiO/VO has to be further investigated, as the enhanced absorption in the observed spectrum behaves differently than the theoretical expectation. To test this hypothesis we employed the ``comb" filter described in \cite{huitson13} and checked for the presence of TiO features. In summary, the filter is a comb of ``in-TiO" and ``out-TiO" bands, which is sensitive to the smaller features within the large TiO signature in the optical region. The same bands were applied to the two isothermal hydrostatic models (with and without TiO/VO) with an equilibrium temperature $T_{eq}=1500\,K$ and surface gravity $10\,m\,s^{-2}$ and theoretical values derived to compare with the differential measurements from our data. Following this approach we found a differential absorption depth, scaled to $H = 414~$km,  of $\Delta z/H = 0.16 \pm 0.69$ from the data and theoretical values for the TiO and non-TiO models were 0.94 and 0.44, respectively. Since the observational result is consistent with no variation, closer to the theoretical value of the non-TiO model, and the overall fit is rather poor, we conclude that the TiO model is not consistent with the data.

An attempt to perform a better fit to the data was also done incorporating isothermal models, pre-computed for the \hat b case (see\,Fig.~\ref{fig:daysidemod}), however with additional parametric adjustments. First,  an unknown ``extra absorber'' at altitude of constant optical opacity $0.03\,cm^2\,g^{-1}$  (from 4000 to 10000~\AA) was incorporated to the isothermal model of \cite{burrows10}. The extra absorber, whatever its nature, increases the ratio of the optical to infrared radii, which is similar to the wavelength dependent radius variation in the \stis and \wfc results and has been used to interpret {\it{Spitzer}} secondary eclipse observations \citep{burrows07b}. In addition, the sodium abundance was increased by a factor of $10^3$ to potentially represent the sodium detection measurement in the spectral bin $5878-5908$~\AA. Decreasing the degrees of freedom with 2, due to the two adjustments in the model, we found that this specific model is an improved fit to the data with $\chi^2 = 105$ and 39 degrees of freedom (see~Fig.~\ref{fig:daysidemod}). The model however, cannot adequately reproduce the enhanced absorption in the bluest and reddest parts of the transmission spectrum (i.e. at decreasing wavelengths shorter than $\lambda \leq 5500$\,\AA\,and longer than $\lambda \geq 8500$\,\AA), as it assumes uniform absorption. Furthermore, even with the enhanced super sodium content the model sodium line prediction seems to be a rather poor match to the observational result. For consistency we also fitted the theoretical models after ignoring the reddest data points of the blue and red transmission spectra. In these cases the  $\chi^2$ decreased significantly down to  66 for 37 degrees of freedom.

Second, the \hat b isothermal model of \cite{fortney08,fortney10} was adjusted with an enhanced (factor of $10^3$ more) Rayleigh scattering in order to reproduce the observational result. This model resulted in $\chi^2 = 122$ and 40 degrees of freedom (see\,Fig.~\ref{fig:daysidemod}). However, incorporating more intense Rayleigh scattering seems to mask/suppress the sodium feature. Removing the two reddest outliers the $\chi^2$  decreased to 66 for 39 degrees of freedom. 

While the list of theoretical models presented in this study is diverse it is far from complete. An optical absorber/scattering may be needed to explain \hat b's optical to near-infrared transmission spectrum, as suggested by the models of  \cite{burrows10} and \cite{fortney08,fortney10} discussed in the last two paragraphs, respectively. However, these two scenarios will be put into future test with further theoretical models attempting to explain our data and to further constrain \hat b planetary atmosphere.

\subsection{The optical blue transmission spectrum}
The G430L transmission spectra derived from a weighted average of individual fits and a simultaneous modelling are found to be in very good agreement (see the lower panel of Fig.~\ref{fig:bluetranspecfig} and Fig.~\ref{fig:simbluetranspecfig}). The main overall property of the transmission spectrum in the range $2900-5700$\,\AA,\ is a relatively smooth featureless slope, analogous to the gradual increase of the absorption with decreasing wavelength in the optical transmission spectrum of \hdtwo~(\citealt{sing08b}, also see \citealt{ballester07, barman07}). Contrary to the high altitude atmospheric haze in \hdone , reported in \cite{pont08} and \cite{ sing11b}, the feature discussed here of \hat b is pronounced in the blue grating only, which significantly differs from the case of HD\,189733\,b, where the feature dominates the complete optical to near-infrared wavelength regime ($3000$\,\AA\,to $10000$\,\AA). Based on the simultaneous fit, the value for the slope is found to be $2100 \pm 672\,km$ over a wavelength range from $2900$\,\AA\,to $5700$\,\AA. \cite{lecavelier08} showed that the effective altitude ($z$) of an exoplanet atmosphere (for a given atmospheric structure and composition) at a wavelength $\lambda$ and assuming a power law for the cross-section of the scattering particles of the form  $\sigma = \sigma_{{\rm{o}}} (\lambda/\lambda_{{\rm{o}}})^\alpha$ (where $\alpha$ is the index for the scattering law) , is given by the following expression:

\begin{equation}
z(\lambda) = \frac{kT}{\mu g} \ln \Bigg( \frac{\xi_{{\rm{abs}}}P_{{\rm{{{\rm{o}}}}}}  }{\tau_{{\rm{eq}}} }   \,\,  \sqrt{\frac{2\pi R_{p}}{kT\mu g}}    \,\,\, \sigma_{{\rm{o}}}  \Big( \frac{\lambda}{\lambda_{{\rm{o}}}}\Big)^{\alpha}      \Bigg) ,
\label{eq:zlambda}
\end{equation}
where $k$ is Boltzmann's constant, $T$ is the atmospheric temperature, $\mu$ is the mean molecular weight of the atmospheric particles, $g$ is the surface gravity, $\xi_{{\rm{abs}}}$ is the abundance of the dominant absorbing species; $P_{{\rm{\sc{{\rm{o}}}}}}$ is the pressure at the reference altitude, $\tau_{{\rm{eq}}}$ is the optical depth at the transmission spectral radius. Following the discussion of \cite{lecavelier08} the slope of the planet radius as a function of the wavelength ($d{z}/d{\ln\lambda}= d{R_{p}}/d{\ln\lambda}$  ) can be derived taking a derivative of $z$ in Equation~\ref{eq:zlambda} with respect to the logarithm of the wavelength\,$\lambda$:

\begin{equation}
\frac{ d{z}}{d{\ln\lambda}} = \alpha \frac{kT}{\mu g} = \alpha H \Rightarrow  \,\,\, \alpha  T = \frac{\mu g}{k} \frac{d R_p}{d \ln \lambda} = \frac{\mu g}{k} \frac{d z}{d \ln \lambda},
\end{equation}
with $H$ representing the atmospheric scale height, see also \cite{howe12}. We find $\alpha T = -7535\pm2411\,K$, assuming $g=877\,cm\,s^2$. The dayside average brightness temperature of \hat b has been determined by \cite{todorov10} to be $T = 1500 \pm 100\,K$, from secondary eclipse observations with {\sl{Spitzer}}. This is likely too high for our terminator spectra at lower temperatures and lower pressures with the water feature detected in \wfc data implying a value closer to $1000\,K$ \citep{wakeford13}. Assuming this temperature, our result translates to $\alpha = -7.5\pm 2.4$. The derived value with uncertainty for $\alpha$ opens up a discussion about the chemical composition of the particles causing the absorption signature within the uncertainty band from $\approx -5$ to $\approx -10$. Generally, and apart from the molecular hydrogen contents, the chemical composition of gases that cause different slopes in the short optical wavelength regimes are not well known.  However, in the case of Rayleigh scattering, the predicted value is $\alpha=-4$, which is close to the lowest border of our result as one of the possibilities. It is hence interesting to speculate the possible source of opacity in the atmosphere of \hat\,b particularly in that case. For example, \cite{lecavelier08} found $\alpha = -4$ given the observational data and the measured slope in the transmission spectrum of HD189733b and suggested MgSiO$_3$ as a possible source of opacity. 

The estimated slope value for \hat b implies a scattering different from Rayleigh, though it depends on the temperature at the different altitudes, and followup observations will be needed to better characterise this slope. 




\cite{ballester07} proposed an alternative explanation for the slope of the transmission spectrum of \hdtwo~below $4000$\,\AA\,as a result of absorption in the Balmer lines and continuum by an optically thin layer of excited hydrogen atoms at high altitude in the planetary atmosphere. Given the single measurement comprising the highest uncertainty of our combined blue transmission spectrum (short ward to 4000 \AA), it is challenging to put this hypothesis into test given the \hat b data. 

A second alternative explanation to the Rayleigh scattering has been proposed by \cite{barman07} who considered an increase of the absorption in the wavelength range  below 5000~\AA~due to atomic lines. The strongest absorption in that wavelength range is expected to be determined mainly by the lines of Ca\,I, Fe\,I, Al\,I and Cr\,I. The strongest among those lines (Ca\,I) is expected to be centred abound $4230$\,\AA. If the abundance of Ca is high one would expect the presence of the Ca\,I and quite possible of some of the other lines in the blue transmission spectrum of \hat b. However, the present spectral bin sizes are much too large (as constrained by the {\it{SNR}} of the data in each spectral bin) to secure a plausible and robust detection of such narrow spectral features.

\subsection{The optical red transmission spectrum}
The G750L low resolution transmission spectrum of \hat b looks marginally flat except for the region around the sodium feature and at wavelength longer than $\lambda \geq 8500$\,\AA. As discussed in Section~\ref{sec:sodiumsec} we detect the core of the sodium (Na\,I) resonance doublet at the $3.3\sigma$ significance level when using a narrow 30\,\AA\,spectral bin, centred at $\lambda=5893 $\,\AA. We searched but did not detect the broad line wings of the sodium feature, which are predicted by both \cite{burrows10} and \cite{howe12} as well as \cite{fortney08,fortney10} cloud free theoretical models with solar abundance (see~Fig.~\ref{fig:sodiumpotassiumfig}). In contrast to the sodium feature the spectrum shows no radius variation around the potassium doublet (K\,I, $\lambda_{{\rm{K\,I}}} = 7665$\,\AA\,and $\lambda_{{\rm{K\,II}}} = 7700$\,\AA ), which is similar to the case of \hdtwo~\citep{charbonneau02, sing08b, jensen11}. Furthermore, the spectrum exhibits a gradual increase of the absorption, with increasing wavelength at wavelengths larger than $8500$\,\AA. 
\cite{huitson13} did not find evidence for this effect in the {\stis} G750L spectral data of WASP-19b, which is part of our survey. Notably, an enhanced absorption is present in the \stis G750L transmission spectrum of \hdtwo~reported in \cite{knutson07}. As \hat b is the second case after \hdtwo~with evidence of enhanced absorption at wavelengths long-ward to $8500$\,\AA~it is likely that the observed effect is real and can be attributed to absorption in these two planetary atmospheres.

\subsubsection{The relative sodium abundance}

Exoplanet transmission spectra allow the abundances to be determined from identified absorption features. In the case of \hat b, when two or more signatures are present, the relative abundance can be measured \citep{lecavelier08, desert09, huitson12}. \hat b is only the second case (after \hdtwo) in the atmosphere of which signatures from both sodium and water are conclusively detected with transmission spectroscopy \citep{charbonneau02, sing12, deming13}. For \hat b, the base of the missing  sodium wings are higher than the peak of the water feature, and a possible explanation could incorporate an enhanced abundance of sodium in the planet atmosphere. 

The relative abundance of the sodium to water, assuming a constant temperature and pressure is given by

\begin{equation}
\frac{\xi_{Na}}{\xi_{H_2O}} = \frac{\sigma_{H_2O}}{\sigma_{Na}} e^{(Z_{{\rm{Na}}}-Z_{{\rm{H_2O}}})/H},
\end{equation}
where $\xi_{Na} $ and $\xi_{H_2O}$ are the abundances, $\sigma_{Na}$ and $\sigma_{H_2O}$ are the wavelength-dependent cross sections at 5893~\AA~and 1.4~$\mu m$, respectively and $Z_{Na}$ and $Z_{H_2O}$ are the altitudes of the observed features. However, in the \hat b case there is further complication as the individual cores of the sodium doublet are not resolved at the spectral resolution of our data and we also did not find evidence for the presence of pressure broadened sodium wings. This limitation prevents determination of the pressure level observed in our transmission spectrum, which is degenerate with absolute abundance measurements. It also hampers the modelling of the sodium lines and limits a derivation of temperature with altitude. The temperature also has a small effect on $\sigma_{{\rm{Na}}}$. If the temperature is not constant as a function of altitude ($z$), the abundance measurement will be affected since higher temperatures cause an increase in $z$ for the same cross-section. 

To estimate the relative sodium to water abundance ratios, ideally one would employ the measured radii of both species at the same altitude, corresponding to similar temperatures. Although we are prevented of such luxury due to the limitation of the low resolution data, we provide an abundance measurement assuming that both features occur in atmospheric regions of the same temperature. Using an average scale height of $414\,km$ (which assumes a temperature of $1000\,K$) and similar temperature between the Na and ${\rm{H_{2}O}}$ features, we estimate an upper limit of the sodium abundance compared to water. We assumed water cross-section $\sigma_{{\rm{H_{2}O}}} = 8.9 \times 10^{-22}\,cm^2$ based on \cite{sharp07}.
To compute the sodium cross section we followed the approach of \cite{huitson12}, however we normalised the integrated cross-section to the stellar spectrum over the $30$\,\AA~bin (used to detect sodium in this study), which  gave $\sigma_{{\rm{Na}}} = 2.3\times10^{-20}\,cm^2$.
Using the derived cross-sections and the measured  $R_{\rm{p}}/R_{\ast} = 0.1223\pm0.0014$ for the sodium line and  $R_{\rm{p}}/R_{\ast}=0.11631\pm0.0003$ for the water feature (see Table~\ref{tab:transpectab}  and \ref{tab:modseltab}, respectively), one finds  $\ln{(\xi_{Na}/\xi_{H_2O})} = +4.6\pm2.2$. For comparison, the solar abundance ratio of sodium to water is $\ln{(\xi_{Na}/\xi_{H_2O})_{solar}} = -5.0$ to $-4.6$ \citep{lodders03, lodders02b, sharp07}.  Despite a large uncertainty, this result indicates an overabundance of sodium compared to water. While we considered an enhanced (factor of $10^3$, i.e. $\ln{(\xi_{Na}/\xi_{H_2O})_{supersolar}}=+2.3$) sodium to water abundance in the specific theoretical model of \cite{burrows10} with the `extra absorber' of uncertain origin, that model was still a rather poor fit to the data. This and additional competing scenarios would be needed to further constrain the atmosphere of \hat b.

\subsection{Interpretation}\label{sec:interpretation}


The complete optical to near-infrared \hat b transmission spectrum from \stis and \wfc constraints several important features of the planet atmosphere. First of all, the red \stis spectra show evidence of sodium ($3.3\sigma$ significance) and lack of potassium. The fact that the sodium cores have been detected but not the wings can potentially be explained by an extra absorber or scatter, which obscurs/masks sodium and potassium features \citep{seager00}. This is in accord with the lack of any evidence in the \stis G750L data of potassium and the overall marginally flat transmission spectrum, which makes the optical spectrum of \hat b similar to that of \hdtwo. This is potentially interesting, because both host stars are of similar spectral type and metallicity. One explanation for the lack of potassium, and presence of sodium could be a general underabundance of potassium in the \hat b atmosphere. Moreover, compared to potassium the sodium atom requires somewhat more energy to ionise. In particular, the first ionisation energies for sodium and potassium are $X_{Na}\,\sim\,500\,{{kJ\,mol^{-1}}}$ and  $X_{K}\,\sim\,420\,{{kJ\,mol^{-1}}}$, respectively, which can partially explain the presence of sodium and no potassium, if there is significant non-thermal ionisation, which is reasonable for highly-irradiated planets. Furthermore, the neutral potassium is $\sim1.7$ times heavier ($m_{a} = 39$ amu\footnote{amu - atomic mass unit}) than the sodium ($m_{a} = 23 $ amu) atom, which may make it harder to vertically mix the species. 

The optical \stis data rules out also an atmosphere dominated by high altitude cloud layer. If it existed, such hypothetical cloud deck would be expected to block (depending on its properties) the complete optical to near-infrared spectrum. Instead, the spectrum exhibits two opposite slopes and a sodium line core in the optical and water feature in the near-infrared.

A comparative analysis of the data with theoretical models of \cite{fortney10} shows that an atmosphere dominated by TiO in the optical regime could hardly explain the optical \stis results. This is no surprise given the medium equilibrium temperature of \hat~is $\sim1000-1300$ K, which is expected not be able to sustain gaseous titanium and vanadium oxides. These compounds are expected to rain down in solids at these temperatures and hence such gases are not expected to determine the opacity in the optical regime. Compared to the water feature in the near-infrared, the \stis optical spectrum has much higher overall opacity. The observed offset cannot be explained as due to stellar activity of the host star nor as a result of the non-linearity of the \wfc detector. A reason for the second case is that the \wfc spectra were obtained well within the linear regime of the camera. Finally, adopting a uniform analysis of the \stis and the \wfc data sets we also rule out a calibration error as a potential explanation for the observed offset. We are hopeful that near-future observations of the \hat b system may further glean more evidence for the difference between the optical and near-infrared transmission spectra. While the theoretical model of \cite{burrows10} with unknown extra absorber and an increased (factor $10^3$) sodium to water abundance ratio is a step in improving the fit to the \hat b \stis and \wfc data, this scenario would be tastable along with additional competing scanarios. If confirmed, it would suggest that the \hat b atmosphere is dominated by an absorber in the optical, partially blocking the sodium wings and completely masking the potassium feature, yet leaving an unmuted water feature at $1.4 \micron$.   

\section{Conclusion}\label{sec:conclusionsec}
In this paper, we report {\hst}/{\stis} transmission spectroscopy of the transiting hot Jupiter \hat\,b, during two transits with the G430L grating ($2900-5700$ \AA) and one transit with G750L grating ($2900-5700$ \AA). The present \stis results are the third (from a large \hst~survey of 8 transiting hot Jupiters) after those of \cite{huitson13} who reported on WASP-19b and  \cite{wakeford13} on \hat b with WFC3. We refine significantly the system physical parameters and \hat b's orbital ephemeris and derive the optical to near-infrared transmission spectrum of the planet. We combine this spectrum with the recently reported  \wfc data by \cite{wakeford13} to constrain the planetary atmospheric properties.

The \stis G750L transmission spectrum is marginally flat, showing evidence for the presence of the sodium (Na\,I) resonance line. We detect the core of the sodium resonance doublet at the $3.3\sigma$ significance level and find no evidence for the presence of sodium pressure-broadened wings. Despite its wider nature, compared to sodium, we find no evidence for the potassium feature in the low resolution \stis data. 

The \stis G430L spectrum shows an increasing absorption shortward to $5300$\,\AA. The measured slope $\alpha=-7.5~\pm~2.4$  (assuming brightness temperature $T=1000\,K$) spans a range of possible scattering signatures approaching Rayleigh (i.e. a slope with $\alpha =-4$) at the lower limit of our measurement. However, more clarification and constraints can be derived with additional followup data, that would increase the accuracy of the derived slope.

Given these characteristics the optical transmission spectrum of \hat b seems analogous to \hdtwo~in many ways. However, a significant difference between both is that \hat b contains only a core of the sodium line, while \hdtwo~shows broad line wings and extra absorption shortward of the sodium central line. 

After a uniform analysis of the white light curves, we combined our \stis spectrum with the spectrum of \citet{wakeford13}, exhibiting the $1.4 \micron$ water signature. The flat optical \stis spectrum is found to show absorption from higher altitudes (at larger planet radius) compared to the \wfc spectrum ($4.3\pm1.6~H$), implying strong optical absorption higher in the atmosphere of \hat b. Further observations are required to confirm this result. Attempting a range of theoretical cloud free atmospheric models of hot Jupiter exoplanets, including these presented in \cite{burrows10} and \cite{howe12} and \cite{fortney08,fortney10}, specifically pre-computed for the system parameters of \hat b, we found poor fits to the data.

The \cite{burrows10} and \cite{howe12} \hat b isothermal model with unknown ``extra absorber" at altitude of constant optical opacity $0.03\,cm^2\,g^{-1}$  (from $4000$ to $10000$\,\AA) and enhanced sodium abundance (by a factor of $10^3$ compared to water) is one scenario that could explain the difference between \stis and \wfc transmission spectrum. However, this model is unable to reproduce the enhanced absorption in the blue and red \stis spectrum. Further modelling incorporating additional scenarios would be needed to test the plausibility of this hypothesis. The planetary atmosphere of \hat b is the second case after \hdtwo~and \hdone~with a conclusive detection of both sodium and water.

\section*{Acknowledgments}
This work is based on observations with the NASA/ESA {\it{Hubble Space Telescope}}. We are grateful to the team at the STScI help desk for their roles in resolving technical issues with the STIS data processing pipeline. N.N. and D.S. acknowledge support from STFC consolidated grant ST/J0016/1. C.H. and P.W. acknowledge a support from STFC grants. All US-based co-authors acknowledge support from the Space Telescope Science Institute under HST-GO-12473 grants to their respective institutions. {\rm{The authors would like to acknowledge the anonymous referee by their useful comments.}}

\appendix{
\section{Model selection for the systematics of the STIS and WFC3 white light curves}
Table~\ref{tab:modseltab}  summarises the results for the relevant competing systematics models, identified using the white light curves of the STIS and WFC3 data sets fitted on individual basis (i.e. not in a joint fit with the remaining data). There are several interesting facts that can be concluded from Table~\ref{tab:modseltab}. First of all, assuming that the photometric uncertainties for each of the three STIS white light curves are dominated by systematics (given the negligible dark current and read-out noise of the STIS CCD detector) the reduced chi square $(\chi^2_{r})$ can be used as in indicator for the level of the photon noise of our data. In particular one can conclude that for the three white light curves the data is at about $\sim44-52\%$ of the theoretical photon noise. Certainly one might doubt that the systematics model could be significantly different from the favoured models in the present analysis. However, the STIS white light curves analysed here show no significant correlations to any of the quantified systematics in order to suggest a functional model for the later. We therefore rely on the lowest BIC values in Table~\ref{tab:modseltab} as a model indicator of a systematics function of the STIS data. Compared to the STIS white light curves the BIC behaves significantly different for the WFC3 data. The statistic constantly decreases with increasing number of the orders of the {\sl{HST}} and planet orbital phases. While we might blindly accept the BIC result and choose model 5 as most appropriate it would be rather hard to point out a physical explanation for the sixth or seventh order {\sl{HST}} orbital phase ($\phi_{t}$). We therefore assume model 3 as a sufficiently good analytic approximation for the systematics in WFC3.

Another interesting fact that can be concluded from Table~\ref{tab:modseltab} is that the individual light curve fits for visits 7, 8 \& 20 as well as the WFC3 show noticeably similar results for the system parameters ($a/R_{\ast}$ and $i$) at the one sigma level, regardless the applied model for the systematics. The later conclusion is also valid for the estimated transit depths (measured with  $R_{\rm{p}}/R_{\ast}$) for both G430L data sets for which the measured $R_{\rm{p}}/R_{\ast}$ values from each visit (separated more than 100 days in time) agree well for a given model. Visit 20 shows one outlier for $R_{\rm{p}}/R_{\ast}$ (model 1, which is unfavored though) which still differs from the highest value at a rather lower level (~1.4 sigma). However, in the result for WFC3 we observe more than $3.5\sigma$ difference between the minimum and maximum values for $R_{\rm{p}}/R_{\ast}$ (models 3 and 1, respectively). This is an important fact which will be discussed again when we paste the STIS and WFC3 transmission spectra in Section~\ref{sec:leveldiffsec}.

\begin{table*}
\caption{Model selection of an appropriate function for the systematics in the white light curves of the STIS and WFC3 data.}
\begin{center}
\begin{tabular}{cccccccccc}
\hline
\hline
Model  &{\sc{BIC}}& $\chi^{2}$&{\sc{DoF}}&{n}&$\chi^2_{r}$ & $i$, $(^{\circ})$ & $a/R_{\ast}$ &   $R_{\rm{p}}/R_{\ast}$    &            $b = a/R_{\ast} \cos{i}$                    \\  
\hline
\multicolumn{10}{l}{\rm{STIS G430L (Visit 7)}} \\
1  &   86   &   47   &   21   &  32 &  2.26    &  $85.773 \pm 0.065$  &  $ 10.006 \pm    0.086$ & $0.11817   \pm  0.00041$ & $0.737 \pm 0.013$  \\
2  &   88   &   47   &   20   &  32 &  2.34    &  $85.798 \pm 0.072$  &  $ 10.037 \pm    0.094$ & $0.11795   \pm  0.00048$ & $0.735 \pm 0.014$  \\   
3  &   92   &   47   &   19   &  32 &  2.45    & $85.806 \pm  0.075$  &  $ 10.049 \pm    0.099$ & $0.11785   \pm  0.00053$ & $0.735 \pm 0.015$  \\
4  &   95   &   46   &   18   &  32 &  2.60    & $85.802 \pm  0.076$  &  $ 10.04   \pm    0.10$    & $0.11786  \pm   0.00053$ & $0.735 \pm 0.015$  \\    
5  &   98   &   46   &   17   &  32 &  2.72    & $85.808 \pm  0.092$  &  $ 10.05    \pm   0.13$    & $0.11787  \pm   0.00061$ & $0.735 \pm 0.019$  \\
6  &   98   &   43   &   16   &  32 &  2.67    & $85.819 \pm  0.079$  &  $ 10.06    \pm   0.10$    & $0.11782  \pm   0.00056$ & $0.734 \pm 0.016$  \\ 
7  & 100   &   45   &   16   &  32 &  2.82    & $85.803 \pm  0.080$  &  $ 10.04   \pm    0.11$    & $0.11783  \pm   0.00054$ & $0.735 \pm 0.016$  \\   
\multicolumn{10}{l}{\rm{STIS G430L (Visit 8)}} \\
1   &    82   &    44   & 20  &  31 &  2.23     &   $85.958  \pm  0.080$  &  $10.32  \pm  0.11$   &   $0.11708  \pm  0.00043$   &   $0.728 \pm 0.016$  \\
2   &    79   &    38   & 19  &  31 &  1.98     &   $85.894  \pm  0.085$  &  $10.24  \pm  0.12$   &   $0.11720  \pm  0.00046$   &   $0.733 \pm 0.017$  \\
3   &    81   &    37   & 18  &  31 &   2.04    &   $85.878  \pm  0.082$   &  $10.22  \pm  0.11$  &   $0.11719  \pm  0.00042$   &   $ 0.734 \pm 0.017$  \\
4  &     83   &    35   & 17  &  31 &   2.05    &   $85.907  \pm  0.088$   &  $10.26  \pm  0.12$  &   $0.11711  \pm  0.00043$   &   $0.732 \pm 0.018$  \\
5  &     84   &    32   & 16  &  31 &   2.00    &   $85.900  \pm  0.086$   &  $10.24  \pm  0.12$  &   $0.11743  \pm  0.00050$   &   $ 0.732 \pm 0.017$  \\
6  &     85   &    30   & 15  &  31 &   2.00    &   $85.936  \pm  0.089$   &  $10.32  \pm  0.13$  &   $0.11703  \pm  0.00054$   &   $0.731 \pm 0.018$  \\
7  &     87   &    32   & 15  &  31 &   2.14    &   $85.900  \pm  0.084$   &  $10.24  \pm  0.12$  &   $0.11743  \pm  0.00048$   &   $0.732 \pm 0.017$  \\
\multicolumn{10}{l}{\rm{STIS G750L (Visit 20)}} \\
1  &    88  &     50    &       21    &    32   &   2.35   &   $85.542   \pm  0.072 $  &   $9.581 \pm 0.077$   &   $0.11783 \pm 0.00052$   &   $0.745 \pm 0.014$  \\
2  &    80  &     39    &       20    &    32   &  1.94    &   $85.495  \pm  0.071$   &   $9.618 \pm 0.076$   &   $0.11843 \pm 0.00051$   &   $0.756 \pm 0.013$  \\
3  &    83  &     38    &       19    &    32   &  2.00    &   $85.488 \pm  0.071$   &   $9.633 \pm 0.078$   &   $0.11844 \pm 0.00053$   &   $0.758 \pm 0.013$  \\    
4  &    83  &     35    &       18    &    32   &  1.94    &   $85.483 \pm  0.072$   &   $9.670 \pm 0.081$   &   $0.11860 \pm 0.00053$   &   $0.762 \pm 0.014$  \\
5  &    87  &     35    &       17    &    32   &   2.03   &   $85.480 \pm  0.079$   &   $9.667 \pm 0.078$   &   $0.11864 \pm 0.00055$   &   $0.762 \pm 0.015$  \\
6  &    88  &     33    &       16    &    32   &   2.03   &   $85.471 \pm  0.075$   &   $9.662 \pm 0.080$   &   $0.11862 \pm 0.00054$   &   $0.763 \pm 0.014$  \\
7  &    89  &     34    &       16    &    32   &   2.10   &   $85.536 \pm  0.083$   &   $9.633 \pm 0.080$   &   $0.11808 \pm 0.00060$   &   $0.750 \pm 0.015$  \\
\multicolumn{10}{l}{\rm{WFC3 G141 (Visit 26)}} \\
1  &  223  &  179  &  70  &  80  & 2.56    &  $85.529  \pm 0.079$   &   $9.784 \pm 0.091$   &   $0.11744 \pm 0.00026$   &   $0.763  \pm 0.015$  \\    
2  &  174  &  125  &  69  &  80  & 1.82    &  $85.599  \pm 0.079$   &   $9.790 \pm 0.089$   &   $0.11723 \pm 0.00023$   &   $0.751  \pm  0.015$  \\
3  &  164  &  111  &  68  &  80  & 1.63    &  $85.676  \pm 0.083$   &   $9.896 \pm 0.095$   &   $0.11631 \pm 0.00030$   &   $0.746   \pm 0.016$  \\  
4  &  164  &  107  &  67  &  80  & 1.60    &  $85.662  \pm 0.083$   &   $9.906 \pm 0.097$   &   $0.11638 \pm 0.00031$   &   $0.749  \pm 0.016$  \\  
5  &  141  &   80   &  66  &  80  & 1.21    &  $ 85.607  \pm 0.083$   &   $9.853 \pm 0.095$   &   $0.11673 \pm 0.00032$   &   $0.755 \pm   0.016$  \\
6  &  144  &   78   &  65  &  80  & 1.20    &  $85.607   \pm 0.083$   &   $9.852 \pm 0.098$   &   $0.11675 \pm 0.00032$   &   $0.754  \pm 0.016$  \\
7  &  145  &   75   &  64  &  80  & 1.17    &  $ 85.598  \pm 0.082$   &   $9.837 \pm 0.096$   &   $0.11680 \pm 0.00032$   &   $0.755  \pm  0.016$  \\
\hline
\multicolumn{7}{l}{Model types for STIS G430L/G750L: }   &      \multicolumn{3}{l}{Model types for WFC3 G141:}  \\
\multicolumn{7}{l}{1 -- $\phi_{t}+\phi^2_{t}+\phi^3_{t}+\phi^4_{t} +t$}                                                        &      \multicolumn{3}{l}{1 -- $\phi_{t}+\phi^2_{t}+\phi^3_{t}+\phi^4_{t}$}       \\
\multicolumn{7}{l}{2 -- $\phi_{t}+\phi^2_{t}+\phi^3_{t}+\phi^4_{t} +t+\omega$}                                        &      \multicolumn{3}{l}{2 -- $\phi_{t}+\phi^2_{t}+\phi^3_{t}+\phi^4_{t} +t$}       \\
\multicolumn{7}{l}{3 -- $\phi_{t}+\phi^2_{t}+\phi^3_{t}+\phi^4_{t} +t+\omega+ x$}                                   &      \multicolumn{3}{l}{3 -- $\phi_{t}+\phi^2_{t}+\phi^3_{t}+\phi^4_{t} +t+t^2$}       \\
\multicolumn{7}{l}{4 -- $\phi_{t}+\phi^2_{t}+\phi^3_{t}+\phi^4_{t} +t+\omega+ x+y$}                               &      \multicolumn{3}{l}{4 -- $\phi_{t}+\phi^2_{t}+\phi^3_{t}+\phi^4_{t}+\phi^5_{t} +t+t^2$}       \\
\multicolumn{7}{l}{5 -- $\phi_{t}+\phi^2_{t}+\phi^3_{t}+\phi^4_{t} +t+\omega+\omega^2+ x+y$}           &      \multicolumn{3}{l}{5 -- $\phi_{t}+\phi^2_{t}+\phi^3_{t}+\phi^4_{t}+\phi^5_{t}+\phi^6_{t}+t+t^2$}       \\
\multicolumn{7}{l}{6 -- $\phi_{t}+\phi^2_{t}+\phi^3_{t}+\phi^4_{t} +t+\omega+\omega^2+ x+x^2+y$}   &      \multicolumn{3}{l}{6 -- $\phi_{t}+\phi^2_{t}+\phi^3_{t}+\phi^4_{t}+\phi^5_{t}+\phi^6_{t}+\phi^7_{t}+t+t^2$}\\
\multicolumn{7}{l}{7 -- $\phi_{t}+\phi^2_{t}+\phi^3_{t}+\phi^4_{t} +t+\omega+\omega^2+ x+y+y^2$}   &      \multicolumn{3}{l}{7 -- $\phi_{t}+\phi^2_{t}+\phi^3_{t}+\phi^4_{t}+\phi^5_{t}+\phi^6_{t} +\phi^7_{t}+\phi^8_{t}+t+t^2$}\\    
\label{tab:modseltab}   
\end{tabular}
\end{center}
\end{table*}
}

\bibliographystyle{mn2e}
\bibliography{nnikolovpaper}

\begin{thebibliography}{78}
\expandafter\ifx\csname natexlab\endcsname\relax\def\natexlab#1{#1}\fi

\bibitem[{{Bakos} {et~al}\mbox{.}(2007){Bakos}, {Noyes}, {Kov{\'a}cs},
  {Latham}, {Sasselov}, {Torres}, {Fischer}, {Stefanik}, {Sato}, {Johnson},
  {P{\'a}l}, {Marcy}, {Butler}, {Esquerdo}, {Stanek}, {L{\'a}z{\'a}r}, {Papp},
  {S{\'a}ri}, \& {Sip{\H o}cz}}]{bakos07}
{Bakos} G.~{\'A}. {et~al.}, 2007, \apj, 656, 552

\bibitem[{{Ballester} {et~al}\mbox{.}(2007){Ballester}, {Sing}, \&
  {Herbert}}]{ballester07}
{Ballester} G.~E., {Sing} D.~K., {Herbert} F., 2007, \nat, 445, 511

\bibitem[{{Barman}(2007)}]{barman07}
{Barman} T., 2007, \apjl, 661, L191

\bibitem[{{B{\'e}ky} {et~al}\mbox{.}(2013){B{\'e}ky}, {Holman}, {Gilliland},
  {Bakos}, {Winn}, {Noyes}, \& {Sasselov}}]{beky13}
{B{\'e}ky} B., {Holman} M.~J., {Gilliland} R.~L., {Bakos} G.~{\'A}., {Winn}
  J.~N., {Noyes} R.~W., {Sasselov} D.~D., 2013, \aj, 145, 166

\bibitem[{{Berta} {et~al}\mbox{.}(2012){Berta}, {Charbonneau}, {D{\'e}sert},
  {Miller-Ricci Kempton}, {McCullough}, {Burke}, {Fortney}, {Irwin}, {Nutzman},
  \& {Homeier}}]{berta12}
{Berta} Z.~K. {et~al.}, 2012, \apj, 747, 35

\bibitem[{{Bevington} \& {Robinson}(2003)}]{bevington03}
{Bevington} P.~R., {Robinson} D.~K., 2003, {Data reduction and error analysis
  for the physical sciences}

\bibitem[{{Brown} {et~al}\mbox{.}(2001){Brown}, {Charbonneau}, {Gilliland},
  {Noyes}, \& {Burrows}}]{brown01}
{Brown} T.~M., {Charbonneau} D., {Gilliland} R.~L., {Noyes} R.~W., {Burrows}
  A., 2001, \apj, 552, 699

\bibitem[{{Burrows} {et~al}\mbox{.}(2001){Burrows}, {Hubbard}, {Lunine}, \&
  {Liebert}}]{burrows01}
{Burrows} A., {Hubbard} W.~B., {Lunine} J.~I., {Liebert} J., 2001, Reviews of
  Modern Physics, 73, 719

\bibitem[{{Burrows} {et~al}\mbox{.}(2007){Burrows}, {Hubeny}, {Budaj},
  {Knutson}, \& {Charbonneau}}]{burrows07b}
{Burrows} A., {Hubeny} I., {Budaj} J., {Knutson} H.~A., {Charbonneau} D., 2007,
  \apjl, 668, L171

\bibitem[{{Burrows} {et~al}\mbox{.}(2010){Burrows}, {Rauscher}, {Spiegel}, \&
  {Menou}}]{burrows10}
{Burrows} A., {Rauscher} E., {Spiegel} D.~S., {Menou} K., 2010, \apj, 719, 341

\bibitem[{{Burrows} \& {Sharp}(1999)}]{burrows99}
{Burrows} A., {Sharp} C.~M., 1999, \apj, 512, 843

\bibitem[{{Charbonneau} {et~al}\mbox{.}(2002){Charbonneau}, {Brown}, {Noyes},
  \& {Gilliland}}]{charbonneau02}
{Charbonneau} D., {Brown} T.~M., {Noyes} R.~W., {Gilliland} R.~L., 2002, \apj,
  568, 377

\bibitem[{{Charbonneau} {et~al}\mbox{.}(2008){Charbonneau}, {Knutson},
  {Barman}, {Allen}, {Mayor}, {Megeath}, {Queloz}, \& {Udry}}]{charbonneau08}
{Charbonneau} D., {Knutson} H.~A., {Barman} T., {Allen} L.~E., {Mayor} M.,
  {Megeath} S.~T., {Queloz} D., {Udry} S., 2008, \apj, 686, 1341

\bibitem[{{Claret}(2000)}]{claret00}
{Claret} A., 2000, \aap, 363, 1081

\bibitem[{{Deming} {et~al}\mbox{.}(2013){Deming}, {Wilkins}, {McCullough},
  {Burrows}, {Fortney}, {Agol}, {Dobbs-Dixon}, {Madhusudhan}, {Crouzet},
  {Desert}, {Gilliland}, {Haynes}, {Knutson}, {Line}, {Magic}, {Mandell},
  {Ranjan}, {Charbonneau}, {Clampin}, {Seager}, \& {Showman}}]{deming13}
{Deming} D. {et~al.}, 2013, \apj, 774, 95

\bibitem[{{D{\'e}sert} {et~al}\mbox{.}(2009){D{\'e}sert}, {Lecavelier des
  Etangs}, {H{\'e}brard}, {Sing}, {Ehrenreich}, {Ferlet}, \&
  {Vidal-Madjar}}]{desert09}
{D{\'e}sert} J.-M., {Lecavelier des Etangs} A., {H{\'e}brard} G., {Sing} D.~K.,
  {Ehrenreich} D., {Ferlet} R., {Vidal-Madjar} A., 2009, \apj, 699, 478

\bibitem[{{Eastman} {et~al}\mbox{.}(2012){Eastman}, {Gaudi}, \&
  {Agol}}]{eastman12}
{Eastman} J., {Gaudi} B.~S., {Agol} E., 2012, {EXOFAST: Fast transit and/or RV
  fitter for single exoplanet}. Astrophysics Source Code Library

\bibitem[{{Fortney} {et~al}\mbox{.}(2008){Fortney}, {Lodders}, {Marley}, \&
  {Freedman}}]{fortney08}
{Fortney} J.~J., {Lodders} K., {Marley} M.~S., {Freedman} R.~S., 2008, \apj,
  678, 1419

\bibitem[{{Fortney} {et~al}\mbox{.}(2010){Fortney}, {Shabram}, {Showman},
  {Lian}, {Freedman}, {Marley}, \& {Lewis}}]{fortney10}
{Fortney} J.~J., {Shabram} M., {Showman} A.~P., {Lian} Y., {Freedman} R.~S.,
  {Marley} M.~S., {Lewis} N.~K., 2010, \apj, 709, 1396

\bibitem[{{Freedman} {et~al}\mbox{.}(2008){Freedman}, {Marley}, \&
  {Lodders}}]{freedman08}
{Freedman} R.~S., {Marley} M.~S., {Lodders} K., 2008, \apjs, 174, 504

\bibitem[{{Gilliland} {et~al}\mbox{.}(1999){Gilliland}, {Goudfrooij}, \&
  {Kimble}}]{gilliland99}
{Gilliland} R.~L., {Goudfrooij} P., {Kimble} R.~A., 1999, \pasp, 111, 1009

\bibitem[{{Goudfrooij} {et~al}\mbox{.}(1998){Goudfrooij}, {Bohlin}, {Walsh}, \&
  {Baum}}]{goudfrooij98a}
{Goudfrooij} P., {Bohlin} R.~C., {Walsh} J.~R., {Baum} S.~A., 1998, {STIS
  Near-IR Fringing. II. Basics and Use of Contemporaneous Flats for
  Spectroscopy of Point Sources (Rev. A)}. Tech. rep.

\bibitem[{{Goudfrooij} \& {Christensen}(1998)}]{goudfrooij98b}
{Goudfrooij} P., {Christensen} J.~A., 1998, {STIS Near-IR Fringing. III. A
  Tutorial on the Use of the IRAF Tasks}. Tech. rep.

\bibitem[{{Grillmair} {et~al}\mbox{.}(2007){Grillmair}, {Charbonneau},
  {Burrows}, {Armus}, {Stauffer}, {Meadows}, {Van Cleve}, \&
  {Levine}}]{grillmair07}
{Grillmair} C.~J., {Charbonneau} D., {Burrows} A., {Armus} L., {Stauffer} J.,
  {Meadows} V., {Van Cleve} J., {Levine} D., 2007, \apjl, 658, L115

\bibitem[{{Hasan} \& {Bely}(1993)}]{hasan93}
{Hasan} H., {Bely} P.~Y., 1993, in Bulletin of the American Astronomical
  Society, Vol.~25, American Astronomical Society Meeting Abstracts, p. 113.06

\bibitem[{{Hasan} \& {Bely}(1994)}]{hasan94}
{Hasan} H., {Bely} P.~Y., 1994, in The Restoration of HST Images and Spectra -
  II, {Hanisch} R.~J., {White} R.~L., eds., p. 157

\bibitem[{{Hayek} {et~al}\mbox{.}(2012){Hayek}, {Sing}, {Pont}, \&
  {Asplund}}]{hayek12}
{Hayek} W., {Sing} D., {Pont} F., {Asplund} M., 2012, \aap, 539, A102

\bibitem[{{Howe} \& {Burrows}(2012)}]{howe12}
{Howe} A.~R., {Burrows} A.~S., 2012, \apj, 756, 176

\bibitem[{{Huitson} {et~al}\mbox{.}(2013){Huitson}, {Sing}, {Pont}, {Fortney},
  {Burrows}, {Wilson}, {Ballester}, {Nikolov}, {Gibson}, {Deming}, {Aigrain},
  {Evans}, {Henry}, {Lecavelier des Etangs}, {Showman}, {Vidal-Madjar}, \&
  {Zahnle}}]{huitson13}
{Huitson} C.~M. {et~al.}, 2013, \mnras, 434, 3252

\bibitem[{{Huitson} {et~al}\mbox{.}(2012){Huitson}, {Sing}, {Vidal-Madjar},
  {Ballester}, {Lecavelier des Etangs}, {D{\'e}sert}, \& {Pont}}]{huitson12}
{Huitson} C.~M., {Sing} D.~K., {Vidal-Madjar} A., {Ballester} G.~E.,
  {Lecavelier des Etangs} A., {D{\'e}sert} J.-M., {Pont} F., 2012, \mnras, 422,
  2477

\bibitem[{{Jensen} {et~al}\mbox{.}(2012){Jensen}, {Redfield}, {Endl},
  {Cochran}, {Koesterke}, \& {Barman}}]{jensen12}
{Jensen} A.~G., {Redfield} S., {Endl} M., {Cochran} W.~D., {Koesterke} L.,
  {Barman} T., 2012, \apj, 751, 86

\bibitem[{{Jensen} {et~al}\mbox{.}(2011){Jensen}, {Redfield}, {Endl},
  {Cochran}, {Koesterke}, \& {Barman}}]{jensen11}
{Jensen} A.~G., {Redfield} S., {Endl} M., {Cochran} W.~D., {Koesterke} L.,
  {Barman} T.~S., 2011, \apj, 743, 203

\bibitem[{{Johnson} {et~al}\mbox{.}(2008){Johnson}, {Winn}, {Narita}, {Enya},
  {Williams}, {Marcy}, {Sato}, {Ohta}, {Taruya}, {Suto}, {Turner}, {Bakos},
  {Butler}, {Vogt}, {Aoki}, {Tamura}, {Yamada}, {Yoshii}, \&
  {Hidas}}]{johnson08}
{Johnson} J.~A. {et~al.}, 2008, \apj, 686, 649

\bibitem[{{Katsanis} \& {McGrath}(1998)}]{katsanis98}
{Katsanis} R.~M., {McGrath} M.~A., 1998, {The Calstis IRAF Calibration Tools
  for STIS Data}. Tech. rep.

\bibitem[{{Knutson} {et~al}\mbox{.}(2008){Knutson}, {Charbonneau}, {Allen},
  {Burrows}, \& {Megeath}}]{knutson08a}
{Knutson} H.~A., {Charbonneau} D., {Allen} L.~E., {Burrows} A., {Megeath}
  S.~T., 2008, \apj, 673, 526

\bibitem[{{Knutson} {et~al}\mbox{.}(2007){Knutson}, {Charbonneau}, {Noyes},
  {Brown}, \& {Gilliland}}]{knutson07}
{Knutson} H.~A., {Charbonneau} D., {Noyes} R.~W., {Brown} T.~M., {Gilliland}
  R.~L., 2007, \apj, 655, 564

\bibitem[{{Knutson} {et~al}\mbox{.}(2010){Knutson}, {Howard}, \&
  {Isaacson}}]{knutson10}
{Knutson} H.~A., {Howard} A.~W., {Isaacson} H., 2010, \apj, 720, 1569

\bibitem[{{Lecavelier Des Etangs} {et~al}\mbox{.}(2008){Lecavelier Des Etangs},
  {Pont}, {Vidal-Madjar}, \& {Sing}}]{lecavelier08}
{Lecavelier Des Etangs} A., {Pont} F., {Vidal-Madjar} A., {Sing} D., 2008,
  \aap, 481, L83

\bibitem[{{Liu} {et~al}\mbox{.}(2008){Liu}, {Burrows}, \& {Ibgui}}]{liu08}
{Liu} X., {Burrows} A., {Ibgui} L., 2008, \apj, 687, 1191

\bibitem[{{Lodders}(1999)}]{lodders99}
{Lodders} K., 1999, \apj, 519, 793

\bibitem[{{Lodders}(2002)}]{lodders02a}
{Lodders} K., 2002, \apj, 577, 974

\bibitem[{{Lodders}(2003)}]{lodders03}
{Lodders} K., 2003, \apj, 591, 1220

\bibitem[{{Lodders}(2009)}]{lodders09}
{Lodders} K., 2009, ArXiv:0910.0811

\bibitem[{{Lodders} \& {Fegley}(2002)}]{lodders02b}
{Lodders} K., {Fegley} B., 2002, \icarus, 155, 393

\bibitem[{{Lodders} \& {Fegley}(2006)}]{lodders06}
{Lodders} K., {Fegley}, Jr. B., 2006, {Chemistry of Low Mass Substellar
  Objects}, p.~1

\bibitem[{{Mandel} \& {Agol}(2002)}]{mandel02}
{Mandel} K., {Agol} E., 2002, \apjl, 580, L171

\bibitem[{{Markwardt}(2009)}]{markwardt09}
{Markwardt} C.~B., 2009, in Astronomical Society of the Pacific Conference
  Series, Vol. 411, Astronomical Data Analysis Software and Systems XVIII,
  {Bohlender} D.~A., {Durand} D., {Dowler} P., eds., p. 251

\bibitem[{{Mayor} \& {Queloz}(1995)}]{mayor95}
{Mayor} M., {Queloz} D., 1995, \nat, 378, 355

\bibitem[{{Narita} {et~al}\mbox{.}(2005){Narita}, {Suto}, {Winn}, {Turner},
  {Aoki}, {Leigh}, {Sato}, {Tamura}, \& {Yamada}}]{narita05}
{Narita} N. {et~al.}, 2005, \pasj, 57, 471

\bibitem[{{Pont} {et~al}\mbox{.}(2008){Pont}, {Knutson}, {Gilliland}, {Moutou},
  \& {Charbonneau}}]{pont08}
{Pont} F., {Knutson} H., {Gilliland} R.~L., {Moutou} C., {Charbonneau} D.,
  2008, \mnras, 385, 109

\bibitem[{{Pont} {et~al}\mbox{.}(2006){Pont}, {Zucker}, \& {Queloz}}]{pont06}
{Pont} F., {Zucker} S., {Queloz} D., 2006, \mnras, 373, 231

\bibitem[{{Redfield} {et~al}\mbox{.}(2008){Redfield}, {Endl}, {Cochran}, \&
  {Koesterke}}]{redfield08}
{Redfield} S., {Endl} M., {Cochran} W.~D., {Koesterke} L., 2008, \apjl, 673,
  L87

\bibitem[{Schwarz(1978)}]{schwarz78}
Schwarz G., 1978, Annals of Statistics, 6, 461

\bibitem[{{Seager}(2011)}]{seager11a}
{Seager} S., 2011, {Exoplanets}

\bibitem[{{Seager} \& {Sasselov}(2000)}]{seager00}
{Seager} S., {Sasselov} D.~D., 2000, \apj, 537, 916

\bibitem[{{Shabram} {et~al}\mbox{.}(2011){Shabram}, {Fortney}, {Greene}, \&
  {Freedman}}]{shabram11}
{Shabram} M., {Fortney} J.~J., {Greene} T.~P., {Freedman} R.~S., 2011, \apj,
  727, 65

\bibitem[{{Sharp} \& {Burrows}(2007)}]{sharp07}
{Sharp} C.~M., {Burrows} A., 2007, \apjs, 168, 140

\bibitem[{{Sing}(2010)}]{sing10}
{Sing} D.~K., 2010, \aap, 510, A21

\bibitem[{{Sing} {et~al}\mbox{.}(2011{\natexlab{a}}){Sing}, {D{\'e}sert},
  {Fortney}, {Lecavelier Des Etangs}, {Ballester}, {Cepa}, {Ehrenreich},
  {L{\'o}pez-Morales}, {Pont}, {Shabram}, \& {Vidal-Madjar}}]{sing11a}
{Sing} D.~K. {et~al.}, 2011{\natexlab{a}}, \aap, 527, A73

\bibitem[{{Sing} {et~al}\mbox{.}(2009){Sing}, {D{\'e}sert}, {Lecavelier Des
  Etangs}, {Ballester}, {Vidal-Madjar}, {Parmentier}, {Hebrard}, \&
  {Henry}}]{sing09b}
{Sing} D.~K., {D{\'e}sert} J.-M., {Lecavelier Des Etangs} A., {Ballester}
  G.~E., {Vidal-Madjar} A., {Parmentier} V., {Hebrard} G., {Henry} G.~W., 2009,
  \aap, 505, 891

\bibitem[{{Sing} {et~al}\mbox{.}(2012){Sing}, {Huitson}, {Lopez-Morales},
  {Pont}, {D{\'e}sert}, {Ehrenreich}, {Wilson}, {Ballester}, {Fortney},
  {Lecavelier des Etangs}, \& {Vidal-Madjar}}]{sing12}
{Sing} D.~K. {et~al.}, 2012, \mnras, 426, 1663

\bibitem[{{Sing} {et~al}\mbox{.}(2011{\natexlab{b}}){Sing}, {Pont}, {Aigrain},
  {Charbonneau}, {D{\'e}sert}, {Gibson}, {Gilliland}, {Hayek}, {Henry},
  {Knutson}, {Lecavelier Des Etangs}, {Mazeh}, \& {Shporer}}]{sing11b}
{Sing} D.~K. {et~al.}, 2011{\natexlab{b}}, \mnras, 416, 1443

\bibitem[{{Sing} {et~al}\mbox{.}(2008{\natexlab{a}}){Sing}, {Vidal-Madjar},
  {D{\'e}sert}, {Lecavelier des Etangs}, \& {Ballester}}]{sing08a}
{Sing} D.~K., {Vidal-Madjar} A., {D{\'e}sert} J.-M., {Lecavelier des Etangs}
  A., {Ballester} G., 2008{\natexlab{a}}, \apj, 686, 658

\bibitem[{{Sing} {et~al}\mbox{.}(2008{\natexlab{b}}){Sing}, {Vidal-Madjar},
  {Lecavelier des Etangs}, {D{\'e}sert}, {Ballester}, \&
  {Ehrenreich}}]{sing08b}
{Sing} D.~K., {Vidal-Madjar} A., {Lecavelier des Etangs} A., {D{\'e}sert}
  J.-M., {Ballester} G., {Ehrenreich} D., 2008{\natexlab{b}}, \apj, 686, 667

\bibitem[{{Snellen} {et~al}\mbox{.}(2008){Snellen}, {Albrecht}, {de Mooij}, \&
  {Le Poole}}]{snellen08}
{Snellen} I.~A.~G., {Albrecht} S., {de Mooij} E.~J.~W., {Le Poole} R.~S., 2008,
  \aap, 487, 357

\bibitem[{{Steele} {et~al}\mbox{.}(2008){Steele}, {Bates}, {Gibson}, {Keenan},
  {Meaburn}, {Mottram}, {Pollacco}, \& {Todd}}]{steele08}
{Steele} I.~A., {Bates} S.~D., {Gibson} N., {Keenan} F., {Meaburn} J.,
  {Mottram} C.~J., {Pollacco} D., {Todd} I., 2008, in Society of Photo-Optical
  Instrumentation Engineers (SPIE) Conference Series, Vol. 7014, Society of
  Photo-Optical Instrumentation Engineers (SPIE) Conference Series

\bibitem[{{Suchkov} \& {Hershey}(1998)}]{suchkov98}
{Suchkov} A., {Hershey} J., 1998, {NICMOS Focus and HST Breathing}. Tech. rep.

\bibitem[{{Todorov} {et~al}\mbox{.}(2010){Todorov}, {Deming}, {Harrington},
  {Stevenson}, {Bowman}, {Nymeyer}, {Fortney}, \& {Bakos}}]{todorov10}
{Todorov} K., {Deming} D., {Harrington} J., {Stevenson} K.~B., {Bowman} W.~C.,
  {Nymeyer} S., {Fortney} J.~J., {Bakos} G.~A., 2010, \apj, 708, 498

\bibitem[{{Torres} {et~al}\mbox{.}(2010){Torres}, {Andersen}, \&
  {Gim{\'e}nez}}]{torres10}
{Torres} G., {Andersen} J., {Gim{\'e}nez} A., 2010, \aapr, 18, 67

\bibitem[{{Torres} {et~al}\mbox{.}(2008){Torres}, {Winn}, \&
  {Holman}}]{torres08}
{Torres} G., {Winn} J.~N., {Holman} M.~J., 2008, \apj, 677, 1324

\bibitem[{{Visscher} {et~al}\mbox{.}(2006){Visscher}, {Lodders}, \&
  {Fegley}}]{visscher06}
{Visscher} C., {Lodders} K., {Fegley}, Jr. B., 2006, \apj, 648, 1181

\bibitem[{{Wakeford} {et~al}\mbox{.}(2013){Wakeford}, {Sing}, {Deming},
  {Gibson}, {Fortney}, {Burrows}, {Ballester}, {Nikolov}, {Aigrain}, {Henry},
  {Knutson}, {Lecavelier des Etangs}, {Pont}, {Showman}, {Vidal-Madjar}, \&
  {Zahnle}}]{wakeford13}
{Wakeford} H.~R. {et~al.}, 2013, ArXiv:1308.2106

\bibitem[{{Winn}(2010)}]{winn10a}
{Winn} J.~N., 2010, ArXiv:1001.2010

\bibitem[{{Winn} {et~al}\mbox{.}(2007){Winn}, {Holman}, {Bakos}, {P{\'a}l},
  {Johnson}, {Williams}, {Shporer}, {Mazeh}, {Fernandez}, {Latham}, \&
  {Gillon}}]{winn07}
{Winn} J.~N. {et~al.}, 2007, \aj, 134, 1707

\bibitem[{{Winn} {et~al}\mbox{.}(2008){Winn}, {Holman}, {Bakos}, {P{\'a}l},
  {Johnson}, {Williams}, {Shporer}, {Mazeh}, {Fernandez}, {Latham}, \&
  {Gillon}}]{winn08}
{Winn} J.~N. {et~al.}, 2008, \aj, 136, 1753

\bibitem[{{Winn} {et~al}\mbox{.}(2009){Winn}, {Holman}, {Henry}, {Torres},
  {Fischer}, {Johnson}, {Marcy}, {Shporer}, \& {Mazeh}}]{winn09}
{Winn} J.~N. {et~al.}, 2009, \apj, 693, 794

\bibitem[{{Wood} {et~al}\mbox{.}(2011){Wood}, {Maxted}, {Smalley}, \&
  {Iro}}]{wood11}
{Wood} P.~L., {Maxted} P.~F.~L., {Smalley} B., {Iro} N., 2011, \mnras, 412,
  2376

\bibitem[{{Zhou} \& {Bayliss}(2012)}]{zhou12}
{Zhou} G., {Bayliss} D.~D.~R., 2012, \mnras, 426, 2483

\end{thebibliography}

\end{document}